\documentclass[12pt,preprint,iop,apj]{emulateapj}  
\usepackage{graphicx}
\usepackage{txfonts}
\newcommand{\ca}{c_{\mathrm{A}}}
\newcommand{\cs}{c_{\mathrm{s}}}

\newcommand{\dtot}{{\rm D}}

\newcommand{\etaA}{\eta_{\rm A}}

\newcommand{\vel}{{\bf v}}
\newcommand{\bmag}{{\bf B}}
\newcommand{\etaC}{\eta_{\rm C}}
\newcommand{\etaCe}{\eta_{\rm C}}

\usepackage{natbib}

\begin{document}

\title{On the spatial scales of  wave heating in the solar chromosphere}

	\shorttitle{Wave heating scales in the chromosphere}

   \author{Roberto Soler$^{1,3}$, Marc Carbonell$^{2,3}$, Jose Luis Ballester$^{1,3}$}

  \affil{$^1$Departament de F\'isica, Universitat de les Illes Balears,
               E-07122, Palma de Mallorca, Spain}
\email{roberto.soler@uib.es}

    \affil{$^2$Departament de Matem\`atiques i Inform\`atica, Universitat de les Illes Balears,
               E-07122, Palma de Mallorca, Spain}

 \affil{$^3$Institute of Applied Computing \& Community Code (IAC$^3$), Universitat de les Illes Balears,
               E-07122, Palma de Mallorca, Spain}

  \begin{abstract}

Dissipation of magnetohydrodynamic (MHD) wave energy  has been proposed as a viable heating mechanism in the solar chromospheric plasma. Here, we use a simplified one-dimensional model of the chromosphere to theoretically investigate the physical processes and the spatial scales that are required for the efficient dissipation of Alfv\'en waves and slow magnetoacoustic waves. We consider the governing equations for a partially ionized hydrogen-helium plasma in the single-fluid MHD approximation and include realistic wave damping mechanisms that may operate in the chromosphere, namely  Ohmic and ambipolar magnetic diffusion, viscosity, thermal conduction, and radiative losses. We perform an analytic local study in the limit of small amplitudes to approximately derive the  lengthscales for critical damping and efficient dissipation of MHD wave energy. We find that the critical dissipation  lengthscale for Alfv\'en waves depends strongly on the magnetic field strength and ranges from 10~m to 1~km for realistic field strengths. The damping of Alfv\'en waves is dominated by Ohmic diffusion for weak magnetic field and low heights in the chromosphere, and by ambipolar diffusion for strong magnetic field and medium/large heights in the chromosphere. { Conversely, the damping of slow magnetoacoustic waves is less efficient, and spatial scales shorter than 10~m are required for critical damping.}  Thermal conduction and viscosity govern the damping of slow magnetoacoustic waves and play an equally important role  at all heights. These results indicate that the spatial scales at which strong wave heating may work in the chromosphere are currently unresolved by observations. 

  \end{abstract}


%________________________________________________________________

\section{INTRODUCTION}

The plasma heating of the upper solar atmosphere is one of the long-standing problems in solar physics. The physical processes responsible for the transport of energy from the solar interior and its dissipation in the atmospheric plasma are under intense research \citep[see, e.g.,][]{2012RSPTA.370.3217P}. One of the mechanisms that has been proposed to explain the transport and dissipation of energy involves the propagation and damping of  magnetohydrodynamic (MHD) waves \citep[see the recent review by][and references therein]{2015RSPTA.37340261A}. Indeed, observations indicate that MHD waves are ubiquitous in the solar atmosphere and can have the required energy to heat the upper layers \citep[see, e.g.,][]{2011Natur.475..463C,2011Natur.475..477M,2014ApJ...795..111H,2015SSRv..tmp...14J}. While the overwhelming presence of the waves is demonstrated by the observations, the physics behind the damping of the waves and the deposition of wave energy into the plasma remains poorly known. 

In this paper, we perform a theoretical study on the conditions for efficient dissipation of MHD wave energy in the solar chromosphere.  The realistic modelling of the physical processes in the chromospheric plasma is challenging. Because of the relatively { low temperature}, the chromospheric plasma is only partially ionized and neutrals are predominant at low and medium altitudes. It has been shown that partial ionization effects  have a strong impact on chromospheric dynamics \citep[see, e.g.,][]{2012ApJ...753..161M,2014SSRv..184..107L}. Ion-neutral collisions may play a crucial role in the release of magnetic energy in the form of heat  \citep{2012ApJ...747...87K}. Therefore, the consideration of partial ionization is a unavoidable requisite for the realistic description of the chromospheric physics.

The role of partial ionization on the damping of Alfv\'en waves has been investigated in detail in the literature. Estimations of the damping rate due to ion-neutral collisions as a function of height in the chromosphere indicate that the damping is most efficient for  waves with high frequencies closer to the local ion-neutral collision frequency \citep[see, e.g.,][among others]{2001ApJ...558..859D,2004A&A...422.1073K,2005A&A...442.1091L,2012A&A...537A..84S}. Recently, \citet{2015A&A...573A..79S} investigated in detail this phenomenon and showed that  high-frequency Alfv\'en waves can be critically damped, i.e., overdamped, because of ion-neutral collisions. The energy carried by these overdamped waves could be efficiently deposited in the  plasma as a result of the strong dissipation. In fact, previous computations of the plasma heating rate obtained from  numerical simulations \citep[see][]{2011ApJ...735...45G,2011JGRA..116.9104S,2013ApJ...777...53T,2013ApJ...765...81R} showed that dissipation of Alfv\'en wave energy can provide a sustained heating over time that is sufficient to compensate the chromospheric radiative losses at low altitudes. So, it is well established that partial ionization effects are important for the correct study of Alfv\'en wave damping and associated heating in the chromosphere.

Concerning magnetoacoustic waves, \citet{2004A&A...422.1073K,2006AdSpR..37..447K} obtained that thermal conduction and viscosity can be important in their damping. Also, results from, e.g., \citet{1994ApJ...435..482P} and \citet{2004A&A...415..739C} suggest that the effect of radiative losses should be taken into account depending on the plasma physical conditions. Conversely, ion-neutral collisions may play a relatively less important role in the direct damping of magnetoacoustic waves \citep[see][]{2007A&A...461..731F}. However, partial ionization should be appropriately accounted for, since the presence of neutrals in the plasma modifies the efficiency of thermal conduction and viscosity. Therefore, as discussed by \citet{2006AdSpR..37..447K}, the correct and complete description of the damping of the various types of MHD waves in the solar atmosphere in general, and in the chromosphere in particular, requires the consideration of all energy dissipation mechanisms, including thermal, collisional, and frictional effects. We add that the correct and complete description of the damping also requires the consideration of all the components of the plasma, namely the various ionized and neutral species.

Here we are concerned about the physical mechanisms that are actually relevant in damping the waves and about the spatial scales that are required for the chromospheric damping to be efficient. We introduce the concept of `critical dissipation lengthscale', which is related to the occurrence of {\em wave cutoffs}. The existence of cutoffs caused by the strong damping of the waves is well documented in the literature \citep[see, e.g.,][among others]{1969ApJ...156..445K,1987ppic.proc..453M,1998ApJ...500..257K,2012A&A...544A.143Z,2013ApJS..209...16S,2013ApJ...767..171S,2014PhPl...21a2110V,2015A&A...573A..79S}. The physical nature of the wave cutoffs resides in the fact that the damping is so strong that the wave restoring force is effectively suppressed \citep[see a more extensive explanation in][]{2013ApJ...767..171S}. As a result, wave perturbations decay in a timescale much shorter than the wave period, a phenomenon called critical damping or overdamping. For practical purposes, this means that the waves are unable to propagate and so they cannot transport energy away from the chromospheric medium.  All the energy initially stored in the wave perturbations is eventually deposited in the plasma in the form of heat. Therefore,  overdamped waves are strong candidates to be connected with efficient plasma heating in the chromosphere \citep{2015A&A...573A..79S}. { Wave cutoffs due to critical damping are physically and conceptually different from  the cutoffs due to plasma stratification studied in many previous works.}

The  purpose of this paper is twofold. On the one hand, we aim to perform a comprehensive theoretical study of the physical mechanisms capable of producing the cutoff of Alfv\'en and magnetoacoustic waves under chromospheric conditions.  It is common among theoretical works that investigate wave damping to focus on the influence of one specific damping mechanism and ignore other effects. We are interested in providing a consistent description of all relevant physical processes that may be involved in strong wave damping. On the other hand, we aim to obtain the critical dissipation lengthscales for the various types of MHD waves. The value of the critical dissipation lengthscale is relevant because it determines the spatial scales needed for  wave heating to become efficient. Strong plasma heating produced by dissipation of MHD wave energy would necessarily require  spatial scales of the order of  the critical dissipation lengthscale.

Another important question from the theoretical point of view is the role of helium in the damping of the waves. The majority of previous works that investigated  wave damping ignored the influence of helium and assumed a chromospheric plasma composed of hydrogen alone. However, it has been shown that the presence of neutral and ionized helium can enhance the damping of Alfv\'en waves due to ion-neutral collisions \citep{2013A&A...549A.113Z}.  A secondary objective of this work is to determine the impact of helium on the efficiency of the damping mechanisms and, consequently, on the critical dissipation lengthscales.

This paper is organized as follows. Section~\ref{sec:basic} contains the description of the chromospheric model, the wave dissipation mechanisms, and the basic equations governing wave propagation and damping. The critical dissipation lengthscales for Alfv\'en waves and slow magnetoacoustic waves in the chromospheric model are investigated in Sections~\ref{sec:firstorder} and \ref{sec:secondorder}, respectively. Later, the results of this article are discussed in Section~\ref{sec:discussion} and some concluding remarks are given in Section~\ref{sec:conclusions}. Finally, the effect of dissipation on the nonlinear coupling of Alfv\'en waves and slow  waves is briefly explored in Appendix~\ref{sec:nonlienar}.

\section{BASIC EQUATIONS}

\label{sec:basic}

\subsection{Partially ionized chromospheric model }

We adopt a one-dimensional, static, gravitationally stratified  model for the chromosphere based on the semi-empirical  model F of \citet{1993ApJ...406..319F}, hereafter the FAL93-F model. The reason for choosing the FAL93-F model instead of more recent models is that the models tabulated in \citet{1993ApJ...406..319F} explicitly provide the variation of the neutral and ionized helium densities with height. The same model was used in the previous work by \citet{2015A&A...573A..79S}, but the influence of helium was not considered there. 

\begin{figure*}
   \centering
  \includegraphics[width=0.99\columnwidth]{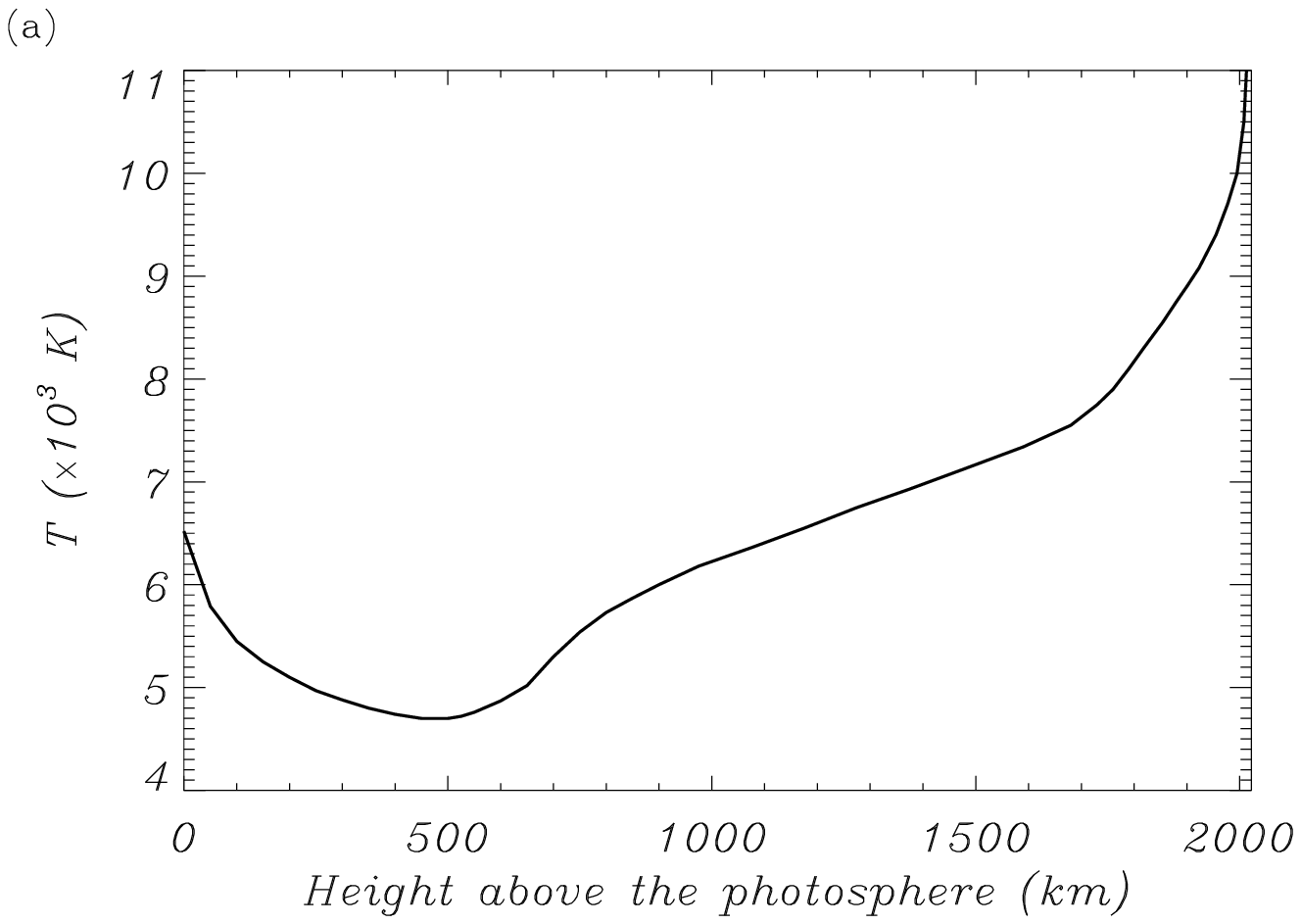}
  \includegraphics[width=0.99\columnwidth]{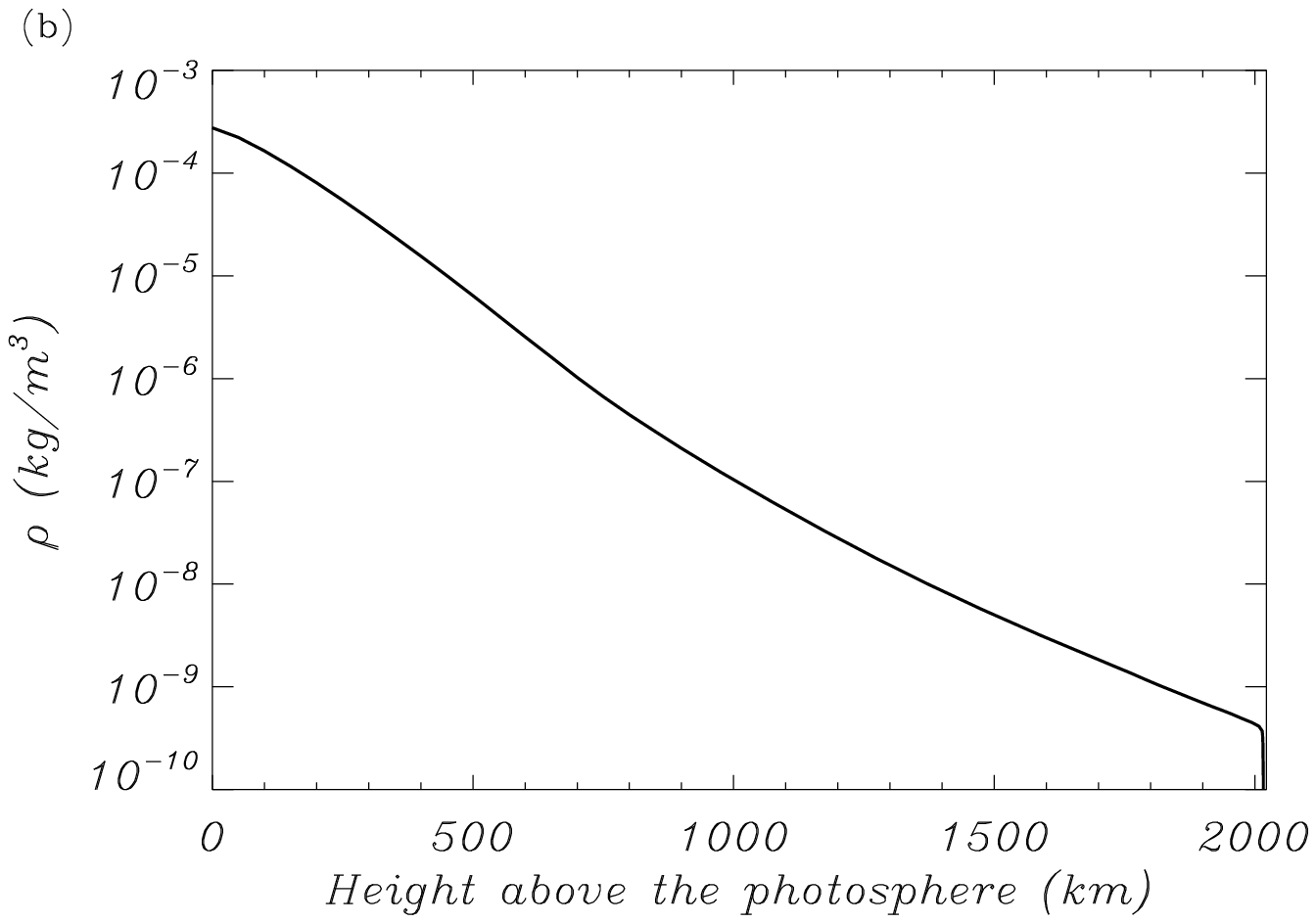}
   \includegraphics[width=0.99\columnwidth]{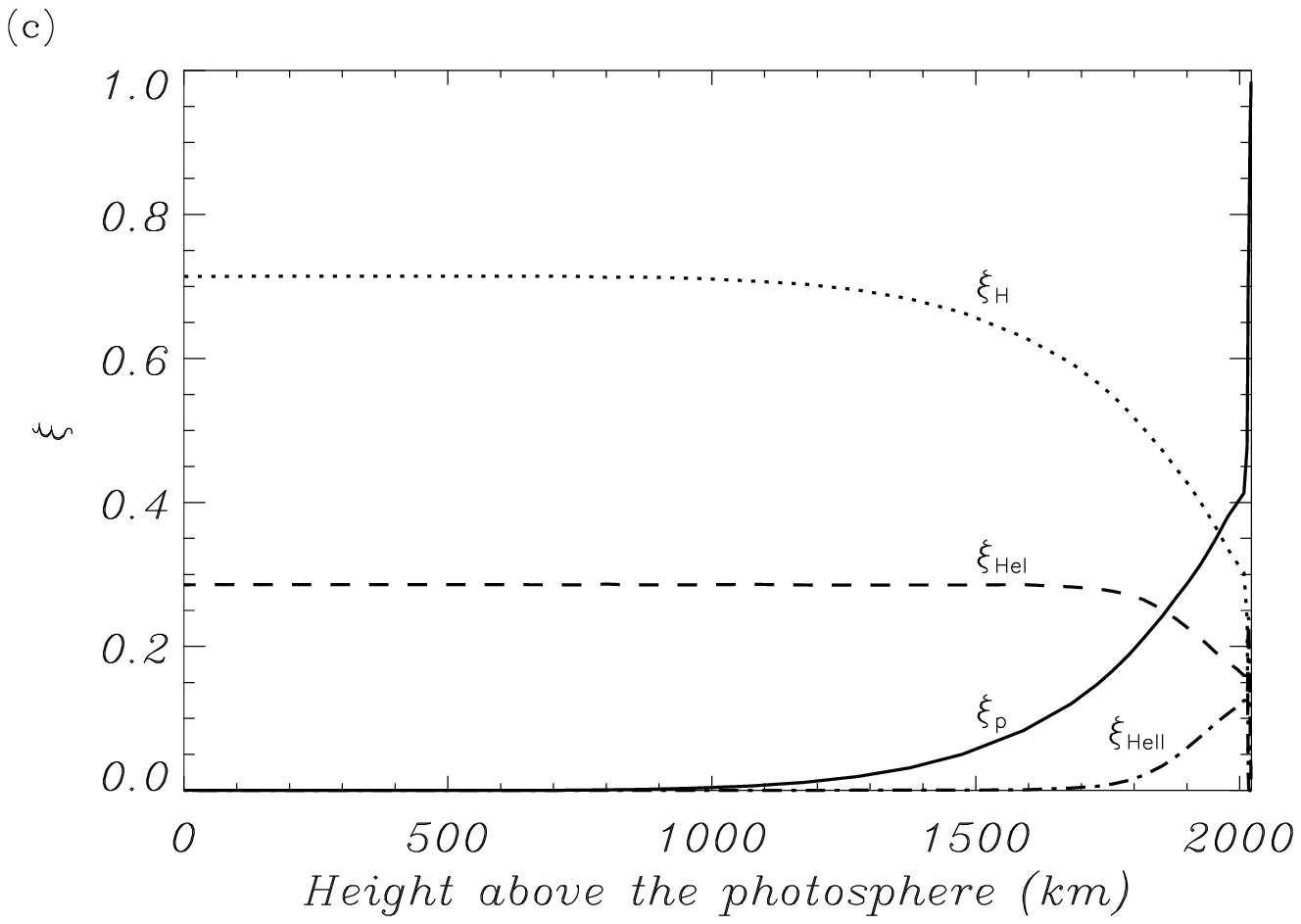}
   \includegraphics[width=0.99\columnwidth]{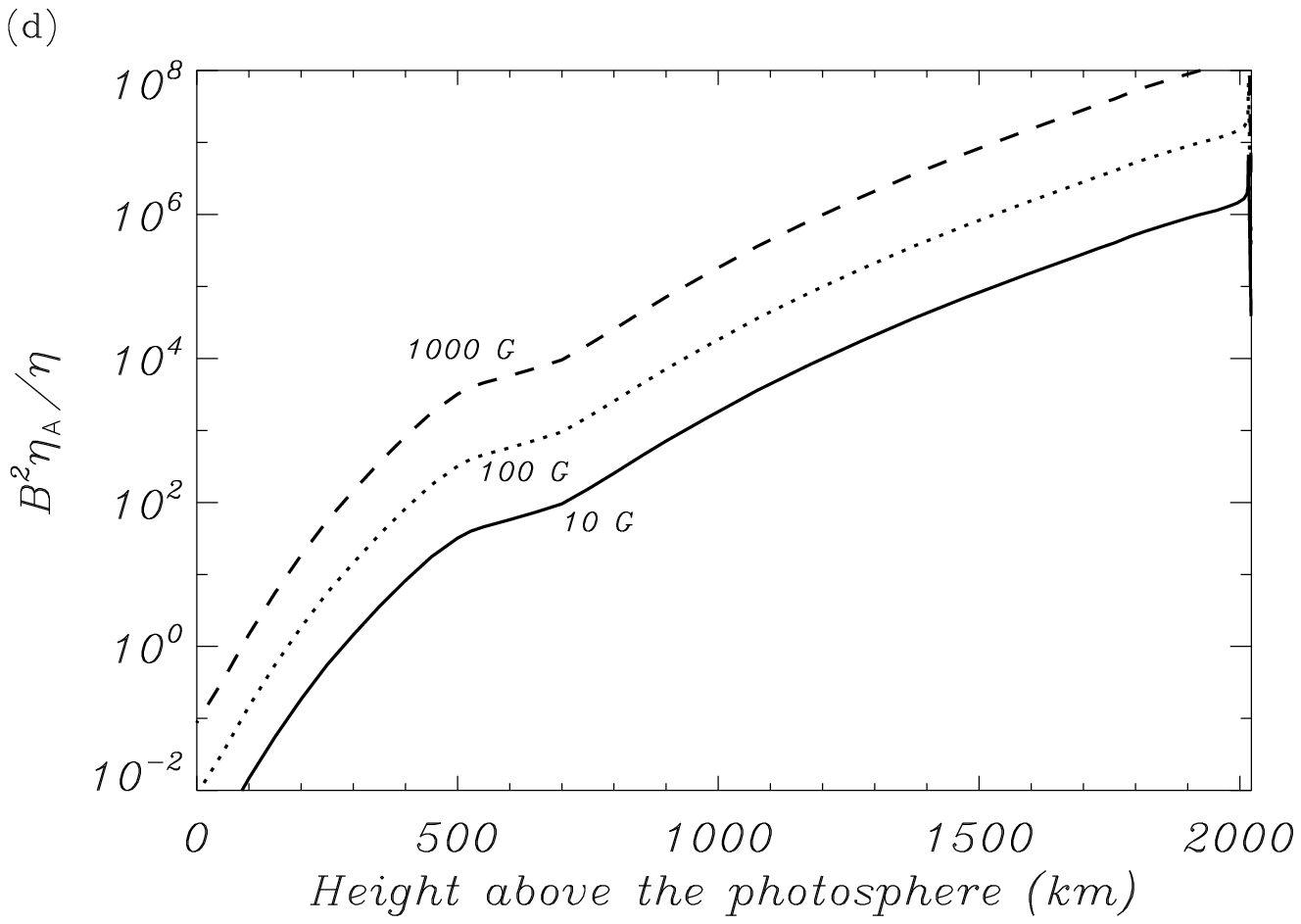}
   \includegraphics[width=0.99\columnwidth]{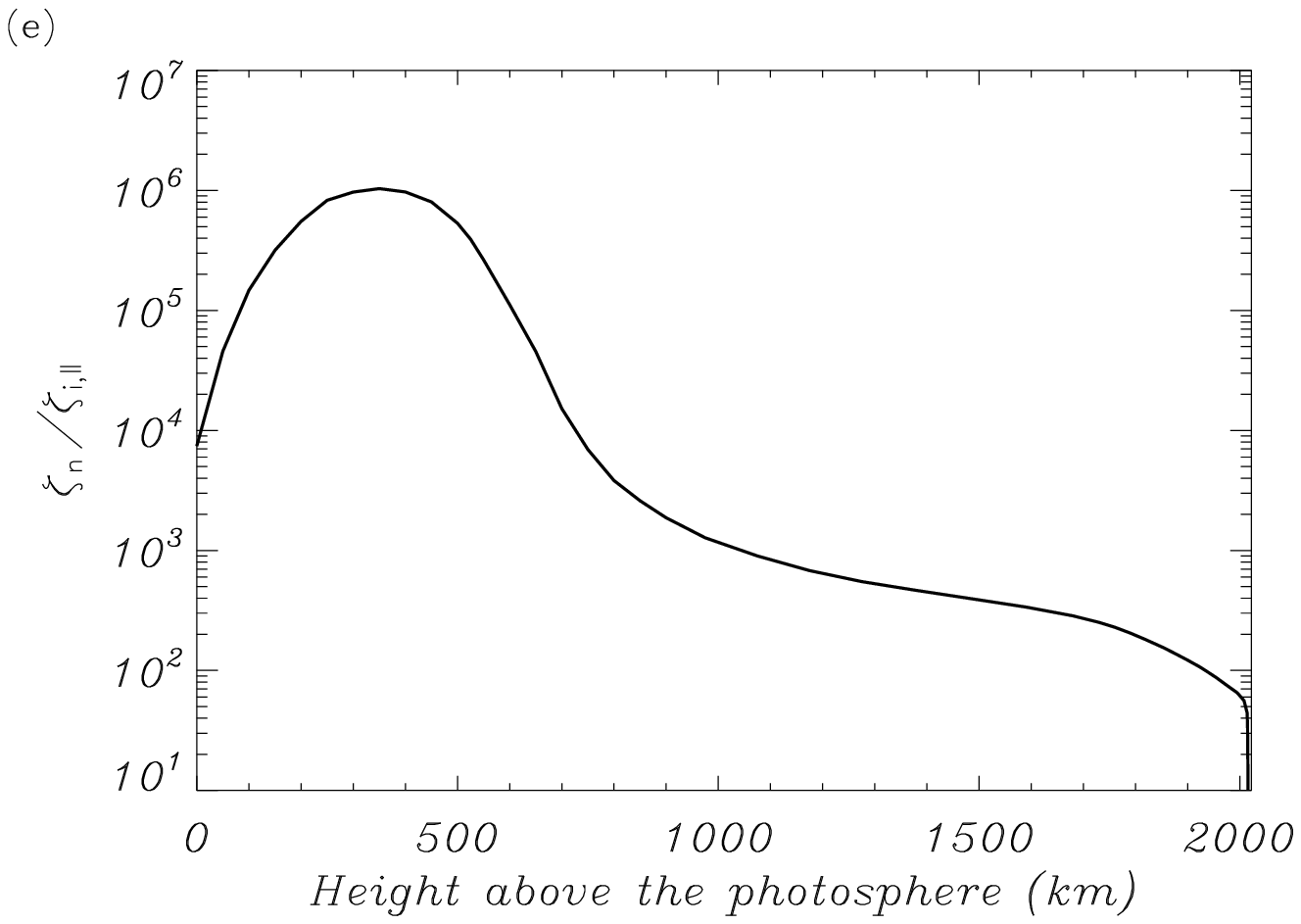}
   \includegraphics[width=0.99\columnwidth]{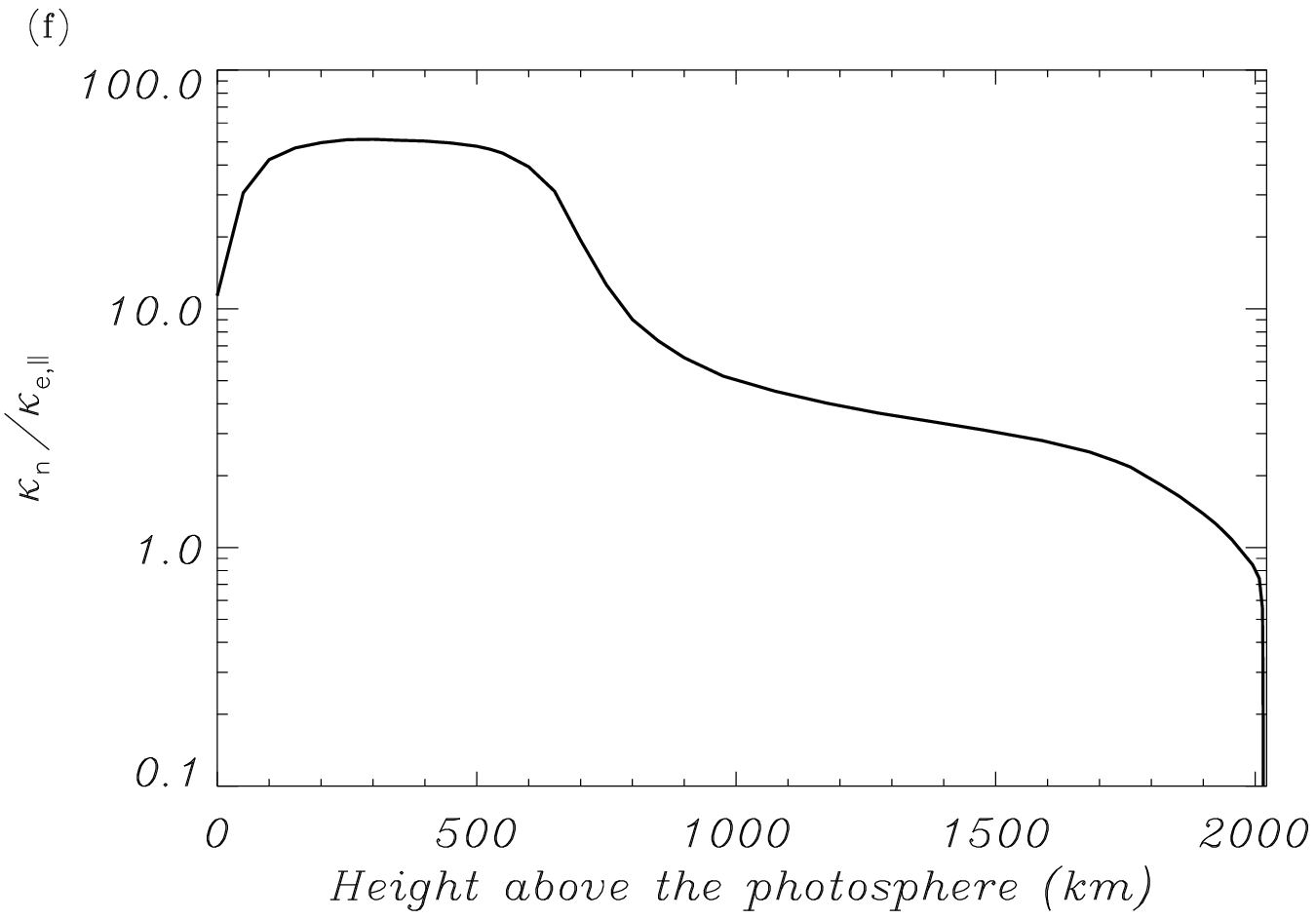}
   \caption{Variation of physical parameters with height above the  photosphere according to the chromospheric  model F of \citet{1993ApJ...406..319F}: (a) temperature, (b) total density, (c) fraction of species, (d) ratio of ambipolar to Ohmic diffusivities for three different magnetic field strengths, (e) ratio of neutral viscosity to  ion parallel viscosity, and (f)  ratio of neutral thermal conductivity to  electron parallel thermal conductivity.}
              \label{fig:xin}%
    \end{figure*}

We treat the chromospheric medium as a  partially ionized hydrogen-helium plasma composed of electrons, protons, neutral hydrogen, neutral helium, and singly ionized helium. Hereafter, subscripts e, p, H, \ion{He}{1}, and \ion{He}{2} explicitly denote these species, respectively.  The presence of doubly ionized helium, \ion{He}{3}, and that of heavier species is ignored because of their negligible abundance in the chromosphere. We define the fraction of species $\beta=$~e, p, H, \ion{He}{1}, or \ion{He}{2} as
\begin{equation}
\xi_\beta = \frac{\rho_{\beta}}{\rho},
\end{equation}
where $\rho_\beta = m_\beta n_\beta$ is the mass density of species $\beta$, with $m_\beta$ and $n_\beta$ the mass particle and number density, and $\rho = \sum_\beta \rho_\beta$ is the total mass density. We note that $\xi_{\rm e} \approx 0$ owing to the very small electron mass. We assume a strong thermal coupling and use the same temperature, $T$, for all the species. Figure~\ref{fig:xin}(a)--(c) shows the dependence on height, $h$, of the temperature, total density, and fraction of the various species, respectively. The sharp chromosphere-to-corona transition region is located at $h\approx$~2,000~km, where the temperature increases and the density decreases abruptly. Hydrogen is mostly neutral for $h \lesssim$~1,500~km,  becomes to be ionized for $h \gtrsim$~1,500~km and is fully ionized for $h\gtrsim$~2,000~km. We note that there is a relatively large abundance of neutral helium at large altitudes in the chromosphere because the temperature is not high enough to fully ionize helium.

The various species in the plasma exchange momentum by means of particle collisions. The strength of the interaction depends on the so-called friction coefficient, $\alpha$. The friction coefficient for collisions between two charged species $q=$~e, p,  or \ion{He}{2} and $q'=$~e, p,  or \ion{He}{2}, is \citep[e.g.,][]{1962pfig.book.....S,1965RvPP....1..205B}
\begin{equation}
\alpha_{qq'} = \frac{n_q n_{q'}e^4 \ln \Lambda_{qq'}}{6\pi\sqrt{2\pi} \epsilon_0^2 m_{q q'} \left( k_{\rm B} T / m_{qq'}\right)^{3/2}}, \label{eq:fric}
\end{equation}
 where  $m_{qq'} = m_q m_{q'}/\left( m_q + m_{q'} \right)$ is the reduced mass, $e$ is the electron charge, $k_{\rm B}$ is Boltzmann's constant,  $\epsilon_0$ is the permittivity of free space,  and $\ln \Lambda_{qq'}$ is Coulomb's logarithm given by \citep[e.g.,][]{1962pfig.book.....S,2013A&A...554A..22V}
\begin{equation}
\ln\Lambda_{\beta\beta'} = \ln \left( \frac{24\pi \epsilon_0^{3/2}  k_{\rm B}^{3/2} T^{3/2}}{e^3 \sqrt{n_{\beta}  +  n_{\beta'}}}   \right).
\end{equation}
In turn, the friction coefficient for collisions between a charged or neutral species $\beta=$~e, p, H, \ion{He}{1}, or \ion{He}{2}, and a neutral species ${\rm n}=$~H or \ion{He}{1} is \citep[e.g.,][]{1965RvPP....1..205B,1970mtnu.book.....C}
\begin{equation}
\alpha_{\beta n} =  n_{\beta} n_{\rm n} m_{\beta \rm n} \left[\frac{8 k_{\rm B}T}{\pi m_{\beta \rm n}}  \right]^{1/2}\sigma_{\rm \beta n}, \label{eq:fricneu}
\end{equation}
where  $\sigma_{\beta \rm  n}$ is the collision cross section. Recent estimations of this parameter can be found in \citet{2013A&A...554A..22V} and an extensive discussion on the importance of its value for wave damping is given in \citet{2015A&A...573A..79S}. 

The expressions of the friction coefficients given above are also valid for self-collisions, i.e., collisions between particles of the same species. The total friction coefficient of species $\beta$ with the other species is
\begin{equation}
\alpha_\beta = \sum_{\beta' \neq \beta} \alpha_{\beta\beta'},
\end{equation}
while the total friction coefficient of species $\beta$ including self-collisions is $\alpha_{\beta,\rm tot} = \alpha_\beta  + \alpha_{\beta\beta}$.
 
\subsection{Single-fluid MHD description of plasma dynamics}

We study the dynamics of the partially ionized chromospheric plasma  within the framework of the single-fluid MHD approximation.  The single-fluid MHD approximation assumes a strong coupling between ions, electrons, and neutrals so that all the species effectively behave as one fluid \citep[see][]{1965RvPP....1..205B}. This approximation is valid in the chromosphere as long as the wave frequencies remain lower than the ion-neutral collision frequency of the plasma { and the wavelengths remain longer than the mean free path of particles between collisions} \citep[see, e.g.,][]{2011A&A...529A..82Z,2013A&A...549A.113Z,2013ApJ...767..171S}.   In this approximation, the basic MHD equations are written in terms of average quantities, while the effects of the interactions, i.e., collisions, between the various species remain in the form of  nonideal terms \citep[see details in, e.g.,][]{2007A&A...461..731F,2012PhPl...19g2508M,2014PhPl...21i2901K}. The basic single-fluid MHD equations used in this work are
\begin{eqnarray}
\frac{\dtot \rho}{\dtot t} &=& -\rho \nabla \cdot \vel, \label{eq:cont} \\
\rho\frac{\dtot \vel}{\dtot t} &=& -\nabla p - \nabla \cdot \hat{\pi} + \rho{\bf g} + \frac{1}{\mu} \left( \nabla \times \bmag \right) \times \bmag, \label{eq:mom} \\
\frac{\partial \bmag}{\partial t} &=& \nabla \times \left( \vel \times \bmag \right) - \nabla \times \left( \eta \nabla \times \bmag \right)\nonumber \\
&& + \nabla \times \left\{ \etaA \left[ \left( \nabla \times \bmag \right) \times \bmag  \right] \times \bmag \right\}, \label{eq:indu} \\
\frac{\dtot p}{\dtot t} &=&  - \gamma p \nabla \cdot \vel + \left( \gamma -1 \right) \mathcal{L}, \label{eq:energy}\\
p&=&  \frac{\rho R T}{\tilde\mu}, \label{eq:state}
\end{eqnarray}
with $\tilde\mu$ the mean atomic weight of a hydrogen-helium plasma given by
\begin{equation}
{\tilde\mu} = \left( 2\xi_{\rm p} + \xi_{\rm H} + \frac{1}{4} \xi_{{\rm He}\,\textsc{i}} + \frac{1}{2} \xi_{{\rm He}\,\textsc{ii}}  \right)^{-1}.
\end{equation}
Equations~(\ref{eq:cont})--(\ref{eq:state}) are the continuity equation, momentum equation, induction equation, energy equation, and the equation of state, respectively. In these equations, $\frac{\dtot}{\dtot t} = \frac{\partial}{\partial t} + \vel \cdot \nabla$ denotes the material or total derivative,  $p$ is the thermal pressure, $\vel$ is the velocity vector, $\bmag$ is the magnetic field vector, $\bf g$ is the acceleration of gravity, $\gamma$ is the adiabatic index, $\mu$ the magnetic permeability, $\hat{\pi}$ is the viscosity tensor, $\eta$ and $\etaA$ are the coefficients of Ohmic and ambipolar diffusion, respectively, $\mathcal{L}$ represents the net effect of all the sources and sinks of energy, and $R$ is the ideal gas constant. We note that Hall's term is ignored in the induction equation due to its irrelevant role in damping the waves \citep{2012A&A...544A.143Z,2015A&A...573A..79S}.

\subsection{Dissipation mechanisms}

Here we discuss the importance of the various dissipation mechanisms that appear in the governing Equations~(\ref{eq:cont})--(\ref{eq:state}). 

Equation~(\ref{eq:indu}) contains two magnetic diffusion terms. The first term on the right-hand side of Equation~(\ref{eq:indu}) is  Ohmic diffusion, which is caused by electron collisions. The coefficient of Ohmic diffusion is given by
\begin{equation}
\eta = \frac{\alpha_{\rm e}}{\mu e^2 n_{\rm e}^2}. \label{eq:eta}
\end{equation}
Equation~(\ref{eq:eta}) corresponds to the classic Spitzer's coefficient of magnetic diffusion \citep{1962pfig.book.....S}. The second term on the right-hand side of Equation~(\ref{eq:indu}) is  ambipolar diffusion. This term contains the effect of ion-neutral collisions. The coefficient of ambipolar diffusion in a hydrogen-helium plasma is given by \citep[see][]{2013A&A...549A.113Z}
\begin{equation}
\etaA = \frac{\xi_{\rm H}^2 \alpha_{{\rm He}\,\textsc{i}} + \xi^2_{{\rm He}\,\textsc{i}}\alpha_{\rm H}  + 2\xi_{\rm H}\xi_{{\rm He}\,\textsc{i}} \alpha_{{\rm H\, He}\,\textsc{i}}}{\mu \left( \alpha_{\rm H} \alpha_{{\rm He}\,\textsc{i}} - \alpha^2_{{\rm H\, He}\,\textsc{i}} \right)}.
\end{equation}
Figure~\ref{fig:xin}(d) displays the ratio $\left| {\bf B} \right|^2\etaA/\eta$ as a function of height in the chromospheric model for three different values of the magnetic field strength. We see that ambipolar diffusion is dominant throughout the chromosphere except at low heights, where Ohmic diffusion can be of importance  for weak magnetic fields. Therefore, we shall consider both Ohmic and ambipolar diffusion in our computations.

Viscosity in a partially ionized plasma is essentially determined  by self-collisions of  ions and  neutrals. Electron viscosity can be safely neglected by virtue of the small electron  mass \citep[see, e.g.,][]{1965RvPP....1..205B, 2012PhPl...19g2508M}.   On the one hand, we assume  the neutral viscosity tensor, $\hat{\pi}_{\rm n}$, to be isotropic because of the absence of the direct effect of the magnetic field on neutrals. Isotropy of neutral viscosity might be altered by ion-neutral collisions \citep[see][]{2014MNRAS.445.1614V}, but  this effect is usually neglected \citep{2012PhPl...19g2508M}. Thus, the neutral viscosity tensor used here is 
\begin{equation}
\hat{\pi}_{\rm n} = - \zeta_{\rm n} \left( \nabla {\bf v} + \left( \nabla {\bf v} \right)^{\top} - \hat{I}\, \frac{2}{3}   \nabla \cdot {\bf v} \right), 
\end{equation} 	
where $\hat{I}$ is the identity tensor and $\zeta_{\rm n}$ is the  coefficient of  neutral viscosity given by
\begin{equation}
 \zeta_{\rm n} = \left( \frac{ m_{\rm H} n^2_{\rm H}}{\alpha_{\rm H,tot}} + \frac{ m_{{\rm He}\,\textsc{i}} n^2_{{\rm He}\,\textsc{i}}}{\alpha_{{\rm He}\,\textsc{i},\rm tot}} \right)  k_{\rm B} T,
\end{equation}
where both neutral hydrogen and neutral helium are included. On the other hand,  the ion viscosity tensor, $\hat{\pi}_{\rm i}$, has a complicated form in  the presence of a magnetic field. It is usually described as the sum of five  components accounting for parallel (or compressive) viscosity, perpendicular (or shear) viscosity, and gyroviscosity \citep[see the full expression in][]{1965RvPP....1..205B}. In the magnetized chromospheric plasma parallel viscosity is several orders of magnitude larger than both perpendicular viscosity and gyroviscosity. Then, we only consider the   ion parallel viscosity component, namely
\begin{equation}
\hat{\pi}_{\rm i}  \approx  -3 \zeta_{\rm i,\parallel} \left( {\bf b b}  - \frac{1}{3} \hat{I} \right) \left( {\bf b b} - \frac{1}{3} \hat{I}  \right) : \nabla {\bf v},
\end{equation}
where ${\bf b}={\bf B}/\left| {\bf B} \right|$ is the unit vector in the direction of the magnetic field and  $ \zeta_{\rm i,\parallel}$ is the coefficient of    ion parallel  viscosity given by
\begin{equation}
\zeta_{\rm i,\parallel} = 0.96 \left( \frac{ m_{\rm p} n^2_{\rm p}}{\alpha_{\rm p,tot}} + \frac{ m_{{\rm He}\,\textsc{ii}} n^2_{{\rm He}\,\textsc{ii}}}{\alpha_{{\rm He}\,\textsc{ii},\rm tot}} \right)  k_{\rm B} T,
\end{equation}
where both protons and singly ionized helium are included. Figure~\ref{fig:xin}(e) displays the ratio $\zeta_{\rm n}/ \zeta_{\rm i,\parallel}$ as a function of height in the chromospheric model. We clearly see that neutral viscosity is several orders of magnitude { more important than} ion viscosity throughout the chromosphere. Because of this result, we are allowed to completely neglect ion viscosity compared to  neutral viscosity. We shall approximate $\hat{\pi} \approx \hat{\pi}_{\rm n}$ in our computations.

The effect of all the sources and sinks of energy is included in the function $\mathcal{L}$ in the right-hand side of the energy equation (Equation~(\ref{eq:energy})). The expression of $\mathcal{L}$ is
\begin{equation}
\mathcal{L} = - \nabla \cdot {\bf q} - L\left(\rho,T  \right) + {\bf j} \cdot {\bf E}^*  - \hat{\pi} : \nabla {\bf v} +H, \label{eq:lenergy}
\end{equation}
where the various terms on the right-hand side are: the divergence of heat flux due to thermal conduction ${\bf q} = -\kappa \nabla T$, with $\kappa$ the thermal conductivity tensor; the radiative loss function $L\left( \rho, T \right)$; the generalized Joule heating ${\bf j} \cdot {\bf E}^*$, { with ${\bf j} = \left(\nabla\times {\bf B}  \right)/\mu$ the current density and ${\bf E}^* = {\bf E} + {\bf v} \times {\bf B}$   the effective electric field}; the viscous heating $\hat{\pi} : \nabla {\bf v}$; and an additional and unspecified  source of  heating $H$.

As happens for viscosity,  thermal conduction  in a magnetized plasma is strongly anisotropic. For convenience, we split the  thermal conductivity tensor into its components parallel, $\kappa_\parallel$, and perpendicular,  $\kappa_\perp$, to the magnetic field direction. In a fully ionized plasma $\kappa_\parallel$ is essentially determined by  electrons, while $\kappa_\perp$ is mainly due to ions and is negligible. In a partially ionized plasma the isotropic { thermal conductivity} of neutrals has to be added to both parallel and perpendicular components.  Thus, the parallel and perpendicular components of { thermal conductivity}  are approximated by $\kappa_\parallel \approx \kappa_{\rm e,\parallel} + \kappa_{\rm n}$ and $\kappa_\perp \approx  \kappa_{\rm n}$, where $\kappa_{\rm e,\parallel}$ and $\kappa_{\rm n}$ are the parallel electron { thermal conductivity}  and the total neutral { thermal conductivity}, respectively, given by
\begin{eqnarray}
\kappa_{\rm e,\parallel} &=& 3.2 \frac{n_{\rm e}^2 k_{\rm B}^2 T}{\alpha_{\rm e,tot}} , \label{eq:kpara} \\
\kappa_{\rm n} & = & \frac{5}{3} \left( \frac{n_{\rm H}^2}{\alpha_{\rm H,tot}} +  \frac{  n^2_{{\rm He}\,\textsc{i}}}{\alpha_{{\rm He},\rm tot}} \right) k_{\rm B}^2 T. \label{eq:kperp}
\end{eqnarray}
Figure~\ref{fig:xin}(e) displays the ratio $\kappa_{\rm n}/ \kappa_{\rm e,\parallel}$ as a function of height in the chromospheric model. The neutral { thermal conductivity}  is found to be more important than the electron { thermal conductivity}  except at large altitudes, where both { thermal conductivities}  are of the same order of magnitude. This is so because hydrogen is largely ionized at those large altitudes, but helium remains mostly neutral. Hence, we shall retain both neutral and electron { thermal conductivities}  in the computations and express the divergence of the heat flux  as
\begin{equation}
\nabla \cdot {\bf q} =  - {\bf B} \cdot \nabla \left( \frac{\kappa_{\rm e,\parallel}}{\left| {\bf B} \right|^2} {\bf B} \cdot \nabla T \right) - \nabla \cdot \left( \kappa_{\rm n} \nabla T \right).
\end{equation}

The radiative loss function, $L\left(\rho,T  \right)$, accounts for plasma cooling  owing to radiative losses. Determining  the chromospheric radiative losses as function of temperature and density is a difficult task that requires complicated numerical solutions of the radiative transfer problem. The radiative loss rate depends, e.g., on the completeness of the atomic model used for the calculation, on the atomic processes included, on the ionization equilibrium, and element abundance assumed, among other factors. The full solution of the radiative transfer problem is beyond the purpose and scope of the present paper. Here we follow a frequent alternative approach to account for the solar plasma radiative losses based on a semi-empirical parametrization of the radiative loss function \citep[see, e.g.,][]{1974SoPh...35..123H,1978ApJ...220..643R,2001ApJ...553..440K,2009A&A...508..751S}. This method enables us to incorporate radiative losses in an approximate way without the need of solving the full radiative transfer problem. The inconvenience of this approach is that the semi-empirical radiative loss function is obtained under the assumption of optically thin plasma, while the cool chromospheric plasmas of interest here do not completely satisfy this condition. Owing to finite optical thickness, the actual radiative losses of the plasma would be probably reduced in some degree compared to the optically thin proxy. This fact should be taken into account when interpreting the effect of radiative losses on the results. The expression of the radiative loss function we use is
\begin{equation}
L\left(\rho,T  \right) = \rho^2 \chi^* T^\alpha, \label{eq:radiationthin}
\end{equation}
where $\chi^*$ and $\alpha$ are piecewise constants depending of the temperature. We use the parametrization of  $\chi^*$ and $\alpha$ given in Table~1 of \citet{2012A&A...540A...7S}, which are obtained from up-to-date computations of radiative losses derived  from the CHIANTI  atomic database \citep{1997A&AS..125..149D,2012ApJ...744...99L} assuming typical abundances in the solar atmosphere. The reader is referred to \citet{2006ApJ...651.1219P} and \citet{2007A&A...469.1109P} for details. 

The generalized Joule term ${\bf j} \cdot {\bf E}^*$ takes into account plasma heating because of dissipation of electric currents. The expression of ${\bf j} \cdot {\bf E}^*$ is
\begin{equation}
{\bf j} \cdot {\bf E}^* = \mu \eta \left| {\bf j}_\parallel \right|^2 + \mu \etaC \left| {\bf j}_\perp \right|^2,
\end{equation}
where $\etaC$ is the so-called Cowling's diffusivity given by
\begin{equation}
\etaC = \eta + \left| \bmag \right|^2 \etaA,
\end{equation}
and ${\bf j}_\parallel $ and ${\bf j}_\perp$ are the components of the current density parallel and perpendicular to the magnetic field, respectively, that can be computed as
\begin{eqnarray}
{\bf j}_\parallel &=& \frac{1}{\mu} \left[ \left( \nabla \times \bmag \right) \cdot {\bf b} \right] {\bf b}, \\
{\bf j}_\perp &=&  \frac{1}{\mu} {\bf b} \times \left[ \left( \nabla \times \bmag \right) \times {\bf b} \right].
\end{eqnarray}
{ Thus, Ohmic magnetic diffusion is caused by the dissipation of parallel currents, while Cowling's magnetic diffusion, i.e., the joint effect of Ohmic and ambipolar diffusion, is caused by the dissipation of perpendicular currents} \citep{2012ApJ...747...87K}. As explained before, we shall retain both Ohmic and ambipolar diffusion in the following computations.

\subsection{Equations for one-dimensional wave propagation}

We  reduce the set of basic equations to the case of the one-dimensional chromospheric model. We use a Cartesian coordinate system and assume that the plasma properties vary along the $x$-direction only, whereas $y$ and $z$ are ignorable directions. Therefore, the $x$-direction corresponds to the vertical direction, so that gravity is oriented in the negative $x$-direction, namely ${\bf g} = (-g, 0,0)$, with $g$ the acceleration of gravity at the solar surface. We can conveniently rotate the  coordinate system for $\vel$ and $\bmag$ to lie on the $xy$-plane with no loss of generality. We redefine the $x$ and $y$ directions as the parallel, $\parallel$, and perpendicular, $\perp$, directions, respectively. Hence, $\vel = (v_\parallel,v_\perp,0)$ and $\bmag=(B_\parallel,B_\perp,0)$. The solenoidal condition $\nabla \cdot \bmag = 0$ imposes that $\partial B_\parallel / \partial x = 0$, while from the $x$-component of Equation~(\ref{eq:indu}) we get $\partial B_\parallel / \partial t = 0$. Therefore, $B_\parallel$ is a constant in both space and time.  In this one-dimensional case, Equations~(\ref{eq:cont})--(\ref{eq:energy}) become
\begin{eqnarray}
\frac{\partial \rho}{\partial t} &=& - \frac{\partial }{\partial x}\left( \rho v_\parallel\right), \label{eq:1d1} \\
\frac{\partial v_\parallel}{\partial t} &=& - v_\parallel \frac{\partial v_\parallel}{\partial x} - \frac{1}{\rho}\frac{\partial p}{\partial x} + \frac{1}{\rho}\frac{\partial }{\partial x}\left( \frac{4\zeta_{\rm n}}{3} \frac{\partial v_\parallel}{\partial x}\right) - g \nonumber \\
&& - \frac{B_\perp}{\mu \rho} \frac{\partial B_\perp}{\partial x}, \label{eq:1d2}  \\
\frac{\partial v_\perp}{\partial t} &=& - v_\parallel \frac{\partial v_\perp}{\partial x} + \frac{1}{\rho}\frac{\partial }{\partial x}\left( \zeta_{\rm n} \frac{\partial v_\perp}{\partial x}\right) + \frac{B_\parallel}{\mu \rho} \frac{\partial B_\perp}{\partial x}, \label{eq:1d3}  \\
\frac{\partial B_\perp}{\partial t} &=& B_\parallel \frac{\partial v_\perp}{\partial x} - \frac{\partial }{\partial x}\left( B_\perp v_\parallel\right) + \frac{\partial}{\partial x}\left(\etaC \frac{\partial B_\perp}{\partial x}\right), \label{eq:1d4}  \\
\frac{\partial p}{\partial t} &=& - v_\parallel \frac{\partial p}{\partial x} - \gamma p \frac{\partial v_\parallel}{\partial x}  + \left(\gamma -1  \right) \left[  \frac{1}{\mu}\etaC \left( \frac{\partial B_\perp}{\partial x} \right)^2  \right. \nonumber \\
&&  +\frac{\partial}{\partial x}\left(\frac{B_\parallel^2 \kappa_{\rm e,\parallel} }{B_\parallel^2 + B_\perp^2}  \frac{\partial T}{\partial x} \right) +\frac{\partial}{\partial x}\left(\kappa_{\rm n} \frac{\partial T}{\partial x} \right)  - L\left(\rho,T  \right)  \nonumber \\
&& \left. + \frac{4\zeta_{\rm n}}{3} \left( \frac{\partial v_\parallel}{\partial x} \right)^2 + \zeta_{\rm n}\left( \frac{\partial v_\perp}{\partial x} \right)^2 + H \right]. \label{eq:1d5} 
\end{eqnarray}
In turn, the variations of temperature are related to those of pressure and density by
\begin{equation}
\frac{1}{T}\frac{\partial T}{\partial t}  = \frac{1}{p} \frac{\partial p}{\partial t} - \frac{1}{\rho} \frac{\partial \rho}{\partial t}. 
\end{equation}

Let us consider the static case so that we set $v_\parallel = v_\perp = B_\perp = 0$ and $\partial/\partial t = 0$. From Equation~(\ref{eq:1d2}) we get
\begin{equation}
\frac{\partial p}{\partial x} = - \rho g. \label{eq:stratified}
\end{equation}
This is the condition for a gravitationally stratified plasma and is  satisfied by the chromospheric model. In turn, from Equation~(\ref{eq:1d5}) we find that the condition for the plasma to be in thermal equilibrium is that the arbitrary heating function must be
\begin{equation}
H = L\left(\rho,T  \right) - \frac{\partial}{\partial x}\left[\left(  \kappa_{\rm e,\parallel}+ \kappa_{\rm n}\right) \frac{\partial T}{\partial x} \right]. \label{eq:heat}
\end{equation}
The arbitrary heating term essentially balances   the background radiative losses since the contribution of thermal conduction, i.e., the second term on the right-hand side of Equation~(\ref{eq:heat}), is almost negligible. We note that this heating term is merely included to formally maintain the semi-empirical chromospheric model in thermal equilibrium. The presence of this arbitrary heating have no effect on the behavior and damping of the waves superimposed on the  background chromosphere.

Inspection of Equations~(\ref{eq:1d1})--(\ref{eq:1d5}) reveals that the velocity longitudinal  component, $v_\parallel$, and the thermal pressure, $p$, are nonlinearly coupled to the velocity and magnetic field transverse components, $v_\perp$ and $B_\perp$. This means that Alfv\'en waves, which are polarized in the transverse direction to the magnetic field, can nonlinearly drive perturbations associated with longitudinally polarized slow magnetoacoustic waves. On the contrary, if $v_\perp$ and $B_\perp$ are initially zero, slow magnetoacoustic waves cannot drive Alfv\'en waves in the plasma. 

\subsection{Approximate local analysis of perturbations}

The purpose of this paper is to obtain expressions for the lengthscales that govern the dissipation of MHD waves in  the chromospheric plasma. To do so, we consider local perturbations superimposed on the background plasma and perform an  approximate study in the limit of small amplitudes.   The use of the local analysis is justified by the fact that, for the waves to be dissipated in the chromosphere, the wavelengths and the associated dissipation lengthscales must be necessarily shorter than the chromospheric gravitational scale height. The pressure scale height in the chromosphere is $\sim 300$~km, so that we must consider shorter lengthscales in this analysis. Consequently, the effect of gravity on the wave perturbations is ignored in the present local analysis, although gravitational stratification is fully taken into account in the background plasma.

We define the dimensionless parameter $\epsilon \equiv \max\left| B_\perp \right| / B_\parallel$ and assume that $\epsilon \ll 1$. Then we write
\begin{eqnarray}
\rho &=& \rho_0 + \epsilon^2 \rho',\\
p &=& p_0 + \epsilon^2 p',\\
T &=& T_0 + \epsilon^2 T',\\
v_\parallel &=& \epsilon^2 v_\parallel', \\
v_\perp &=& \epsilon v_\perp',\\
B_\parallel &=& B_0,\\
B_\perp &=& \epsilon B_\perp',
\end{eqnarray}
where the subscript 0 denotes a background quantity and the prime $'$ denotes a perturbation. To separate the perturbations of Alfv\'en waves from those of slow waves, the perturbations of the perpendicular components of velocity and magnetic field are assumed to be first-order in $\epsilon$, while the perturbations of the remaining  quantities are assumed to be second order in $\epsilon$. We substitute these quantities in Equations~(\ref{eq:1d1})--(\ref{eq:1d5}) and separate the various terms according to their order with respect to $\epsilon$.  

The conditions for static equilibrium  (Equations~(\ref{eq:stratified}) and (\ref{eq:heat})) are consistently recovered from the zeroth-order terms in $\epsilon$. The first-order equations in $\epsilon$ govern the behavior of $v_\perp$ and $B_\perp$, and so  they describe linear Alfv\'en waves, namely
\begin{eqnarray}
\frac{\partial v_\perp'}{\partial t} &=& \tilde{\zeta}_{\rm n} \frac{\partial^2 v_\perp'}{\partial x^2}+  \frac{B_0}{\mu \rho_0} \frac{\partial B_\perp'}{\partial x},  \label{eq:al1} \\
\frac{\partial B_\perp'}{\partial t} &=&  B_0 \frac{\partial v_\perp'}{\partial x} +\etaCe \frac{\partial^2 B_\perp'}{\partial x^2}, \label{eq:al2}
\end{eqnarray}
where $\tilde{\zeta}_{\rm n}=\zeta_{\rm n}/\rho_0$. The second-order equations in $\epsilon$  involve $v_\parallel'$, $p'$, $\rho'$, and $T'$, so that they describe linear slow magnetoacoustic waves and also the generation of slow magnetoacoustic waves due to the nonlinear coupling with the Alfv\'en waves, namely
\begin{eqnarray}
\frac{\partial \rho'}{\partial t} &=&  - \rho_0 \frac{\partial v_\parallel'}{\partial x}, \label{eq:second1} \\
\rho_0 \frac{\partial v_\parallel'}{\partial t} &=& -\frac{\partial p'}{\partial x} + \frac{4\zeta_{\rm n}}{3}\frac{\partial^2 v_\parallel'}{\partial x^2}  - \frac{1}{2\mu} \frac{\partial {B_\perp'}^2}{\partial x}, \label{eq:second2} \\
\frac{\partial p'}{\partial t} &=& - \gamma p_0 \frac{\partial v_\parallel'}{\partial x} + \left(\gamma -1  \right) \left[ \left( \kappa_{\rm e,\parallel} + \kappa_{\rm n} \right) \frac{\partial^2 T'}{\partial x^2} -\rho_0  L_\rho \rho'  \right. \nonumber \\
&& \left. - \rho_0  L_T T'  + \frac{1}{\mu}\etaCe \left(\frac{\partial {B_\perp'}}{\partial x}\right)^2  + \zeta_{\rm n} \left(\frac{\partial {v_\perp'}}{\partial x}\right)^2 \right], \label{eq:second3}
\end{eqnarray}
where $L_\rho$ and $L_T$ are the partial derivatives of the radiative loss function with respect to density and temperature, respectively,  namely
\begin{equation}
L_\rho \equiv  \left(\frac{\partial L}{\partial \rho}\right)_{\rho_0,T_0}, \qquad L_T \equiv  \left(\frac{\partial L}{\partial T}\right)_{\rho_0,T_0}.
\end{equation}
In addition, $p'$, $\rho'$, and $T'$ are related through  the equation of state as
\begin{equation}
\frac{p'}{p_0} = \frac{\rho'}{\rho_0} + \frac{T'}{T_0}. \label{eq:state2}
\end{equation}
The equations for higher orders in $\epsilon$ represent nonlinear corrections on the Alfv\'en and slow waves. In this approximate study we consider sufficiently low amplitudes for the high-order corrections in $\epsilon$ to be negligible. Therefore we restrict ourselves to the first-order and second-order equations in $\epsilon$.

\section{CRITICAL DISSIPATION LENGTHSCALE OF ALFV\'EN WAVES}
\label{sec:firstorder}

\begin{figure*}
   \centering
  \includegraphics[width=0.99\columnwidth]{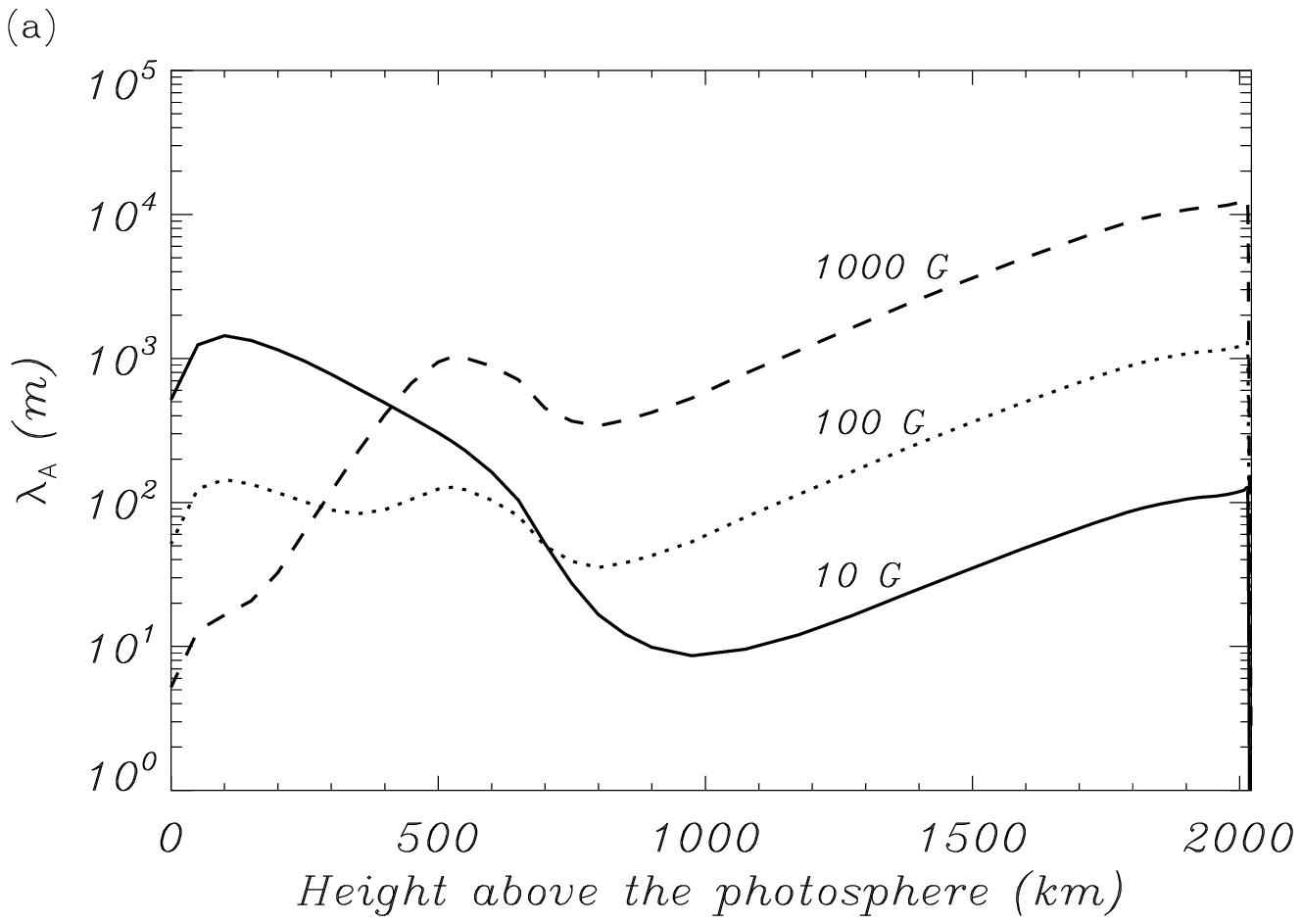}
   \includegraphics[width=0.99\columnwidth]{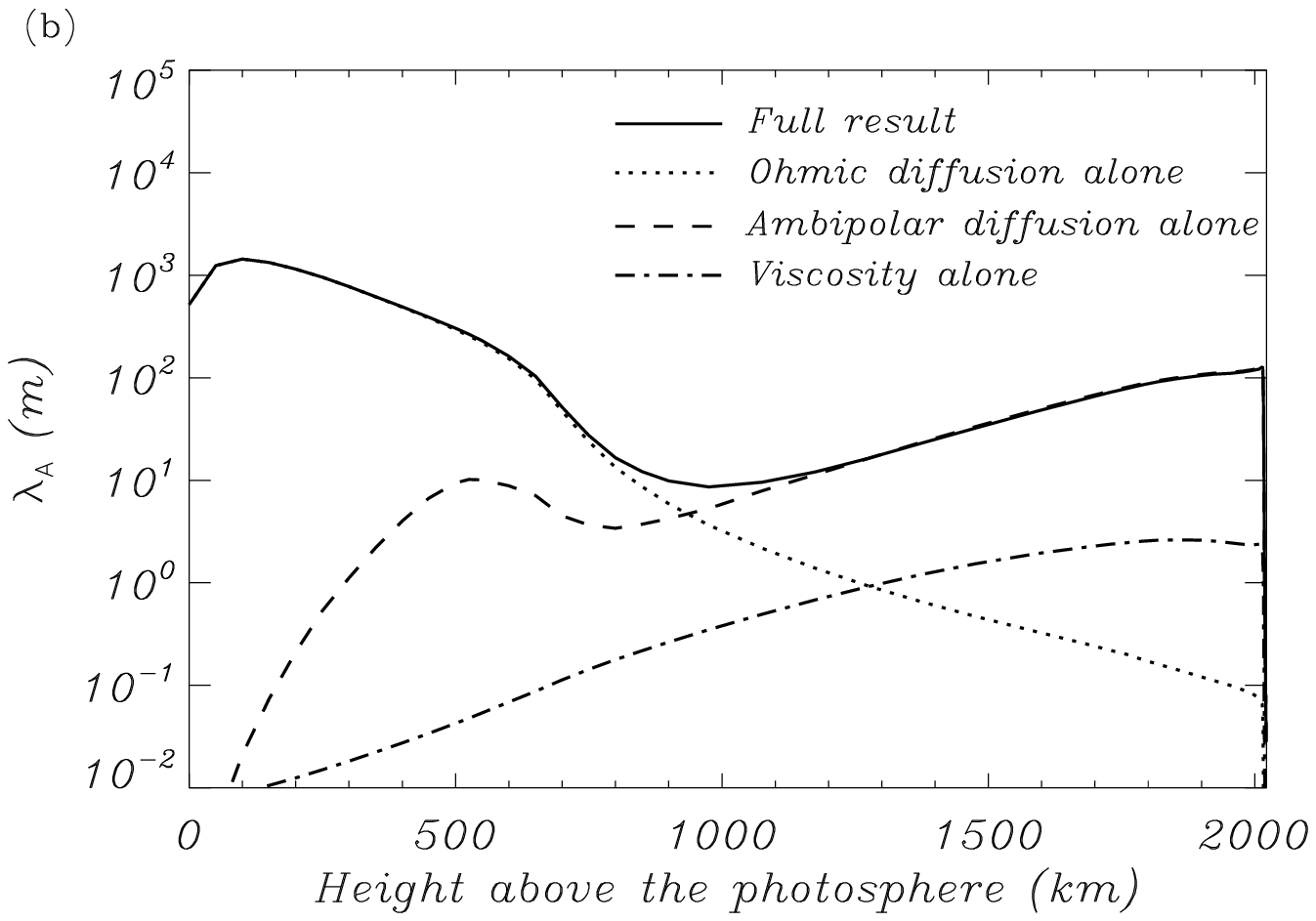}
     \includegraphics[width=0.99\columnwidth]{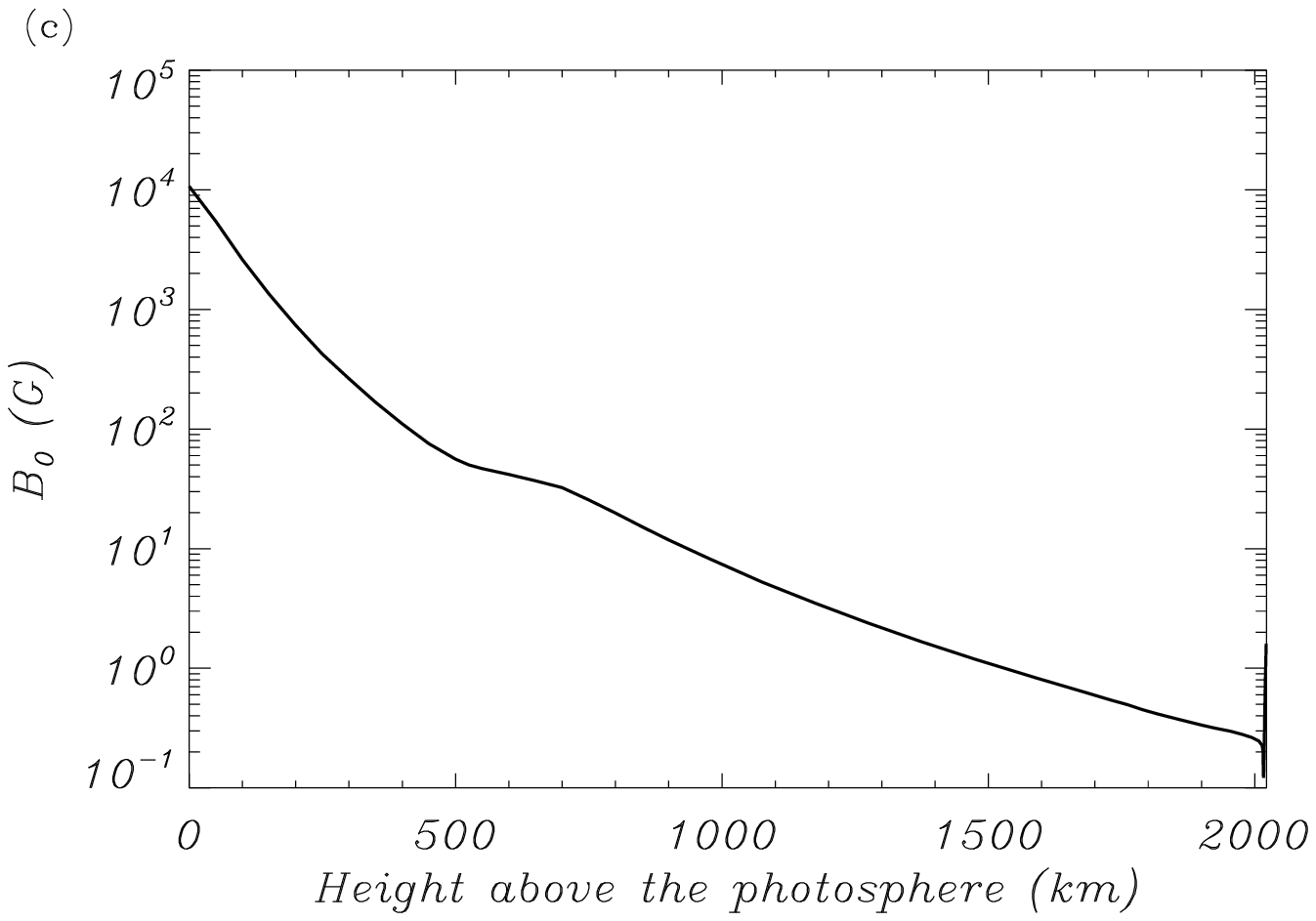}
   \includegraphics[width=0.99\columnwidth]{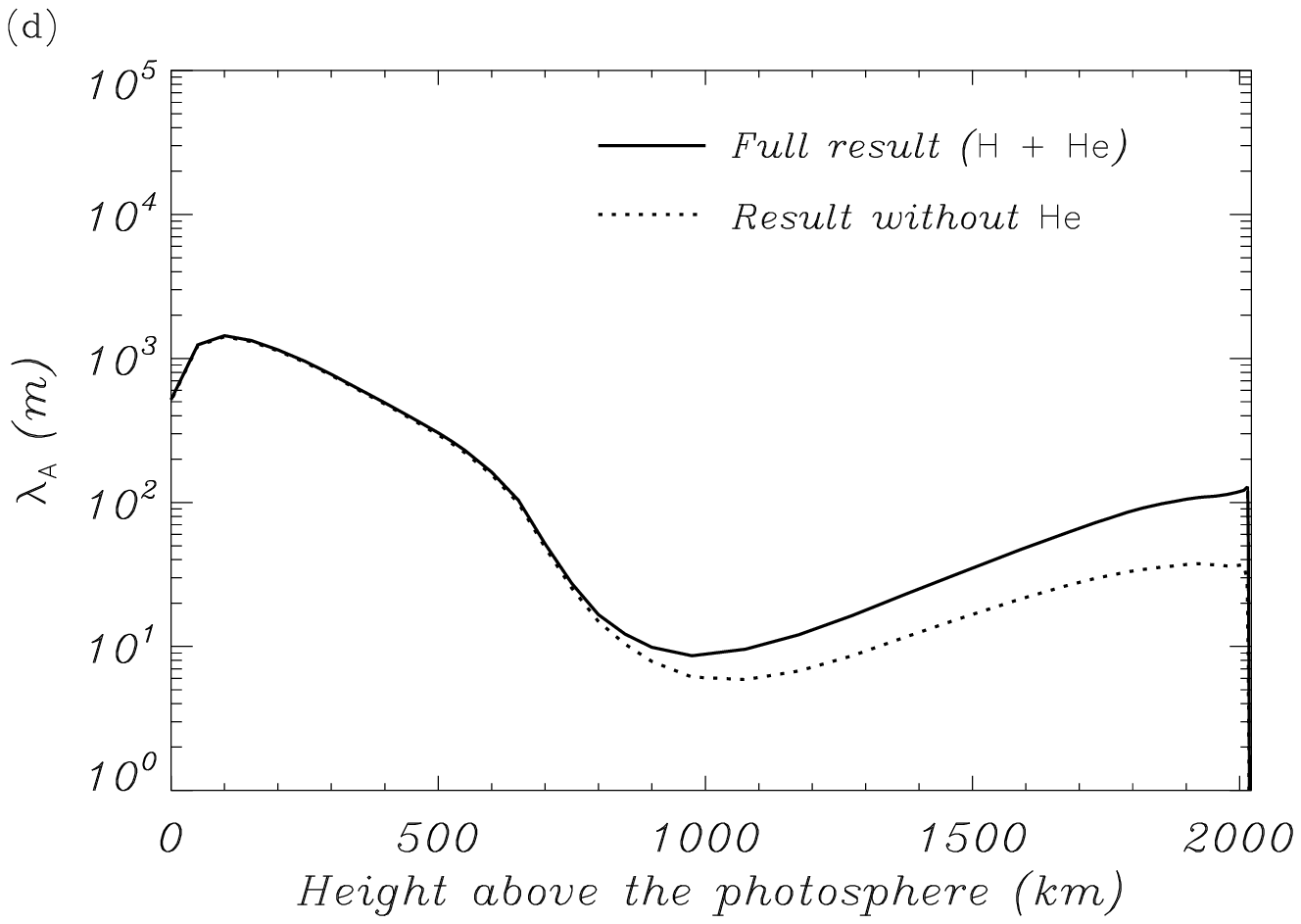}
   \caption{(a) Critical dissipation lengthscale for Alfv\'en waves as a function of height above the solar photosphere according to the chromospheric model F of \citet{1993ApJ...406..319F}. The various linestyles correspond to a magnetic field strength, $B_0$, of 10~G (solid), 100~G (dotted), and 1000~G (dashed). (b) Effect of the various damping mechanisms on the critical dissipation lengthscale for Alfv\'en waves with $B_0 = 10$~G: full result (solid), effect of Ohmic diffusion alone (dotted), effect of ambipolar diffusion alone (dashed), and effect of viscosity alone (dot-dashed). (c) Threshold magnetic field strength for which ambipolar diffusion dominates over Ohmic diffusion as a function of height. (d) Effect of helium on the critical dissipation lengthscale for Alfv\'en waves with $B_0 = 10$~G: full result (solid) and result without helium (dotted). }
              \label{fig:lamcrit}%
    \end{figure*}

To start with, we consider Alfv\'en waves, which are governed by Equations~(\ref{eq:al1}) and (\ref{eq:al2}). These two equations involve the transverse components of $\vel$ and $\bmag$ and can be appropriately combined into an equation for $B_\perp'$ only, namely
\begin{equation}
\frac{\partial^2 B_\perp'}{\partial t^2} -  \ca^2 \frac{\partial^2 B_\perp'}{\partial x^2} -\left( \etaCe + \tilde{\zeta}_{\rm n} \right) \frac{\partial^3 B_\perp'}{\partial x^2 \partial t} + \etaCe \tilde{\zeta}_{\rm n}\frac{\partial^4 B_\perp'}{\partial x^4} = 0, \label{eq:alflin}
\end{equation}
where $\ca^2 = B_0^2 / \mu \rho_0$ is the  Alfv\'en velocity squared. Equation~(\ref{eq:alflin}) governs linear Alfv\'en waves damped by Cowling's diffusion and viscosity. It can be shown that $v_\perp'$ satisfies the same equation. 

Let us assume that a packet of Alfv\'en waves is generated impulsively by a perturbation with a certain spatial form. Any disturbance in the plasma can be expressed as a superposition of Fourier modes. Each Fourier mode is characterized by a wavenumber, $k$, and a frequency, $\omega(k)$. Thus, we  write
\begin{equation}
B_\perp' = \sum_k \mathcal{B}_k \exp\left[ i k x - i\omega(k) t \right], \label{eq:fourierby}\label{eq:fourierby}
\end{equation}
 where $\mathcal{B}_k$ is the amplitude of the Fourier mode with wavenumber $k$. We substitute this expression into Equation~(\ref{eq:alflin}) to find the dispersion relation for small-amplitude, linear Alfv\'en waves in a viscous-resistive plasma, namely
\begin{equation}
\omega^2(k) + i  k^2 \left( \etaCe + \tilde{\zeta}_{\rm n} \right) \omega(k) - k^2 \ca^2 -k^4  \etaCe \tilde{\zeta}_{\rm n} = 0. \label{eq:alfreldisper}
\end{equation}
Equation~(\ref{eq:alfreldisper}) was previously derived by, e.g., \citet{1961hhs..book.....C} and \citet{2012A&A...544A.143Z}.

 The individual Fourier modes can be studied separately since no interaction between different Fourier modes of the wavepacket takes place in the small-amplitude, linear regime considered here. We focus on one individual Fourier mode of the wavepacket, so we set $k$  to a fixed value and $\omega(k)$ is obtained by solving Equation~(\ref{eq:alfreldisper}). The analytic solution to  Equation~(\ref{eq:alfreldisper}) is
\begin{equation}
\omega(k) = \pm k \ca \left( 1 - \frac{k^2\left( \etaCe - \tilde{\zeta}_{\rm n} \right)^2}{4\ca^2} \right)^{1/2} - i \frac{k^2}{2}\left( \etaCe + \tilde{\zeta}_{\rm n} \right). \label{eq:omegaimp}
\end{equation}
 The first term on the right-hand side of Equation~(\ref{eq:omegaimp}) is the real part of the frequency, namely $\omega_{\rm R}(k)$, and the second term is the imaginary part, namely $\omega_{\rm I}(k)$. The $+$ and $-$ signs in front of $\omega_{\rm R}(k)$ correspond to waves propagating towards the positive $x$-direction (upwards) and towards the negative $x$-direction (downwards), respectively. Hence, the initial perturbation naturally splits into two wavepackets propagating in opposite directions. The imaginary part of $\omega(k)$ defines a damping time scale for the Fourier mode, so that its amplitude decreases in time following an exponential law, namely $\exp \left( -  \left| \omega_{\rm I}(k) \right| t \right)$. Ideal undamped Alfv\'en waves are recovered when no dissipation is present ($\etaCe=\tilde{\zeta}_{\rm n}=0$), so that  $\omega_{\rm R}(k) =  \pm k \ca$ and $\omega_{\rm I}(k) = 0$ in that case. However, in the presence of Cowling's diffusion and/or viscosity Alfv\'en waves  are damped.

The quality factor, $Q(k)$, is a dimensionless parameter that characterizes how efficiently damped a specific Fourier mode is. The quality factor is commonly defined as
\begin{equation}
Q(k) \equiv \frac{1}{2}  \left|\frac{\omega_{\rm R}(k)}{\omega_{\rm I}(k)}\right|. \label{eq:quality}
\end{equation}
The quality factor is independent of the direction of wave propagation. The strength of wave damping depends on the value of $Q(k)$. According to the definition of $Q(k)$ given in Equation~(\ref{eq:quality}), the waves are weakly damped or {\em underdamped} when $Q(k) > 1/2$. In that case, most of the wave energy would not dissipate in the chromosphere. The larger $Q(k)$, the less efficient damping, so that no wave energy dissipation takes place if $Q(k)\to\infty$. Conversely, when $Q(k) < 1/2$  the waves are {\em overdamped} and the wave energy dissipation is very strong: most of the wave energy  is dissipated. The most extreme situation takes place when $Q(k) =0$, which corresponds to a  {\em wave cutoff}. In a cutoff scenario,  waves cannot propagate  and all the energy stored in the waves is deposited in the plasma. The expression of $Q(k)$ computed from the Alfv\'en wave frequency (Equation~(\ref{eq:omegaimp})) is
\begin{equation}
Q(k) = \left\{ \begin{array}{lll}
\frac{\left| \etaCe - \tilde{\zeta}_{\rm n} \right|}{\etaCe + \tilde{\zeta}_{\rm n} }\frac{\sqrt{k_{\rm A}^2 - k^2}}{2k}  & {\rm if} & k < k_{\rm A}, \\
0 & {\rm if} & k \geq k_{\rm A},  
\end{array} \right.
\end{equation}
where $k_{\rm A}$ is a  cutoff wavenumber for Alfv\'en waves given by
\begin{equation}
k_{\rm A} = \frac{2\ca}{\left| \etaCe - \tilde{\zeta}_{\rm n} \right|}. \label{eq:cutoff}
\end{equation}
The Fourier modes of the wavepacket with $k \geq k_{\rm A}$ cannot propagate in the form of Alfv\'en waves since their quality factor is zero. As discussed before, the physical reason for this result is that damping is extremely efficient for these large-$k$ modes, so that their propagation is inhibited due to the very strong energy dissipation. Thus, the  wavepackets that would eventually propagate away from the chromosphere necessarily contain Fourier modes with $k < k_{\rm A}$ alone, while the Fourier modes with $k \ge k_{\rm A}$ would be completely dissipated in the chromospheric plasma.

The cutoff wavenumber defines the lengthscale for which dissipation is total, namely 
\begin{equation}
\lambda_{\rm A} = \frac{2\pi}{k_{\rm A}} = \frac{\pi \left| \etaCe - \tilde{\zeta}_{\rm n} \right|}{\ca}. \label{eq:alfvencritical}
\end{equation}
The critical dissipation lengthscale is plotted in Figure~\ref{fig:lamcrit}(a) as a function of height for three values of the magnetic field strength, namely $B_0=$~10, 100, and 1000~G. In the lower chromosphere, the larger the magnetic field strength, the shorter the critical dissipation lengthscale. The behavior is the opposite one in the middle and upper chromosphere, where the critical dissipation lengthscale increases when the magnetic field strength is increased. The reason for this result resides in the mechanisms responsible for the damping of the waves. This is explored in Figure~\ref{fig:lamcrit}(b), which shows the critical dissipation lengthscale for $B_0 = 10$~G computed when only one specific damping mechanism is retained. We recall that Cowling's diffusivity, $\etaCe$, contains both Ohmic, $\eta$, and ambipolar, $\eta_{\rm A}$, diffusivities. It can be seen that Ohmic diffusion dominates the damping at low heights, while ambipolar diffusion becomes the most important mechanism at large heights. Viscosity plays no important role throughout the chromosphere. The switch from Ohmic damping to ambipolar damping depends on the value of the magnetic field strength. This is shown in Figure~\ref{fig:lamcrit}(c), where the threshold magnetic field strength is plotted as a function of height. This  threshold magnetic field strength is computed as $B_0 \sim \sqrt{\eta/\eta_{\rm A}}$. We see that strong fields of the order of kG are required at the low chromosphere for ambipolar diffusion to be more important than Ohmic diffusion. However, only a few tens of G are needed in the medium chromosphere, and very weak fields are needed in the upper chromosphere. These results concerning the relative importance of Ohmic and ambipolar diffusion are consistent with the expected importance of the two effects according to Figure~\ref{fig:xin}(d).

We have also explored the particular effect of helium on the value of the critical dissipation lengthscale. Figure~\ref{fig:lamcrit}(d) compares the Alfv\'en wave critical dissipation lengthscale for $B_0 = 10$~G when helium is taken into account and when the effect of helium is neglected in the dissipation mechanisms. This figure indicates that the effect of helium is not important when the damping is caused by Ohmic diffusion. For the magnetic field strength used in Figure~\ref{fig:lamcrit}(d), this happens when $h \lesssim 900$~km. However, helium should be taken into account when the damping is dominated by ambipolar diffusion. In  Figure~\ref{fig:lamcrit}(d), this happens when $h \gtrsim 900$~km. The reason for the important impact of helium on the ambipolar damping is that helium remains largely neutral in the chromosphere. Obviously, since the importance of helium is linked to that of ambipolar diffusion, the particular height at which the effect of helium becomes important depends on the value of the magnetic field strength considered (see again Figure~\ref{fig:lamcrit}(c)). In general, the influence of helium should not be neglected \citep[see also][]{2013A&A...549A.113Z}.

\section{CRITICAL DISSIPATION LENGTHSCALE OF SLOW MAGNETOACOUSTIC WAVES}
\label{sec:secondorder}

 \begin{figure*}
   \centering
  \includegraphics[width=0.99\columnwidth]{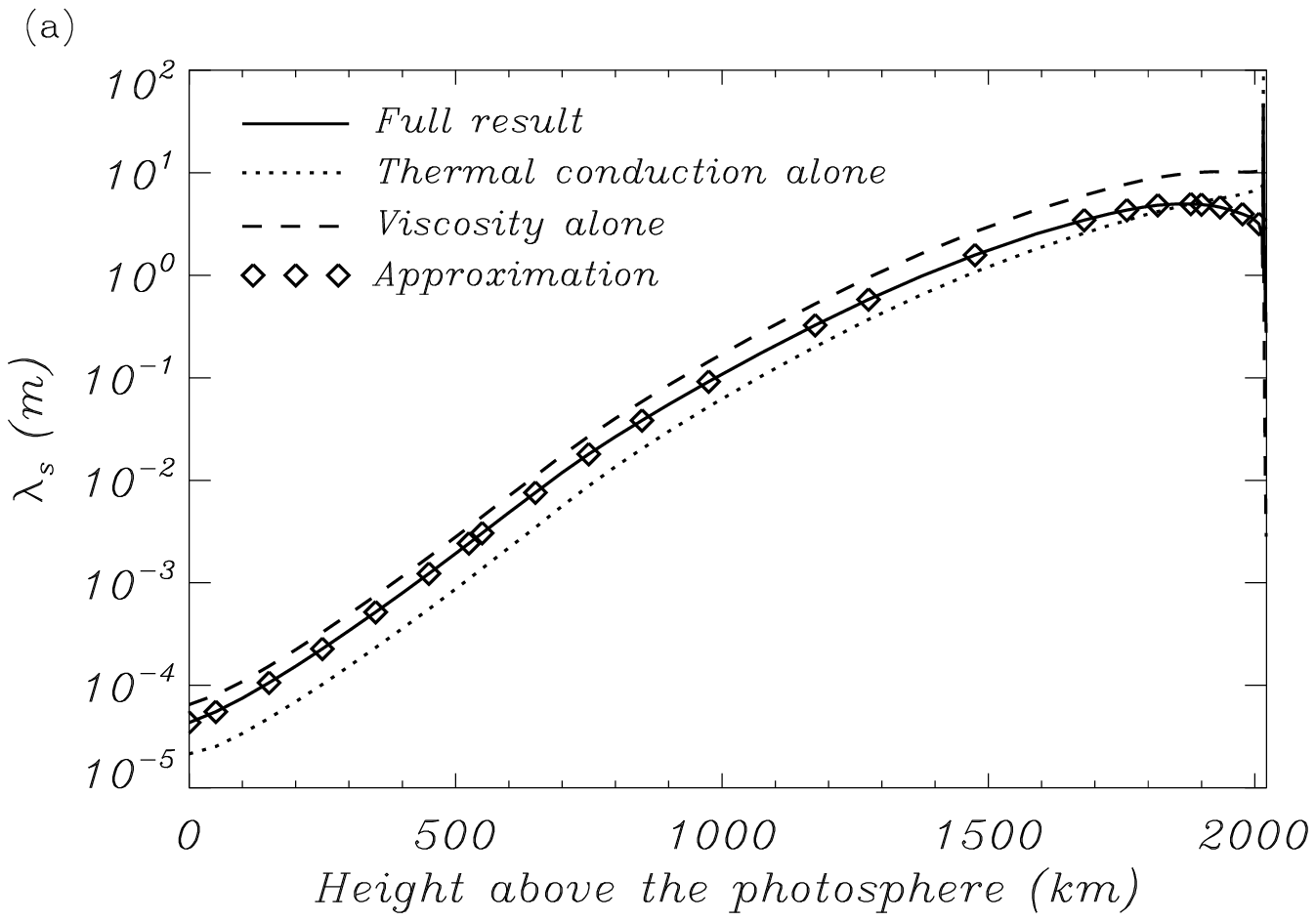}
   \includegraphics[width=0.99\columnwidth]{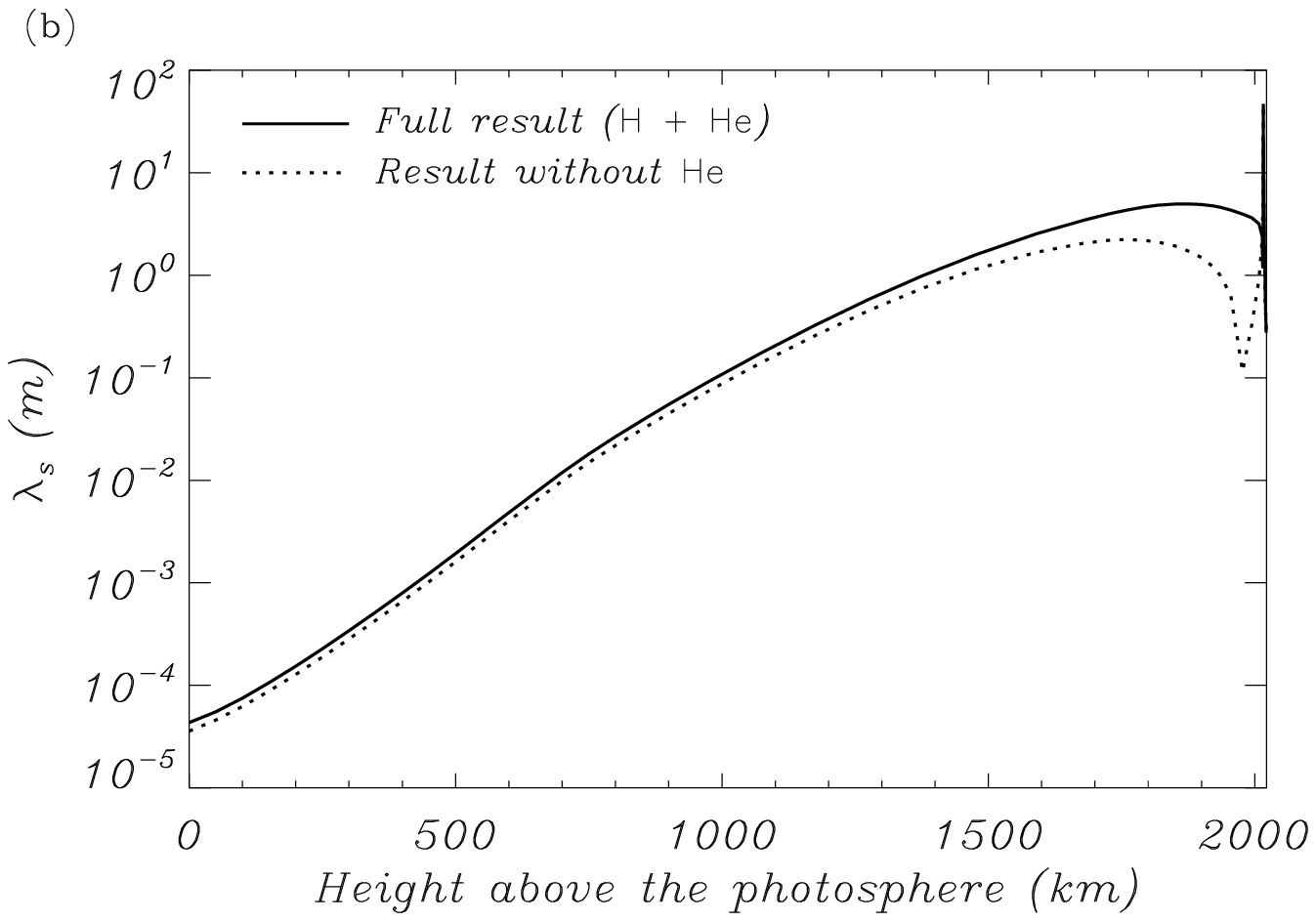}
   \caption{(a) Critical dissipation lengthscale for slow magnetoacoustic waves as a function of height above the solar photosphere according to the chromospheric model F of \citet{1993ApJ...406..319F}. The various linestyles correspond to the full result (solid), the effect of thermal conduction alone (dotted), the effect of viscosity alone (dashed), and the analytic approximation given in  Equation~(\ref{eq:slowlength}) (symbols). (b) Effect of helium on the critical dissipation lengthscale for slow magnetoacoustic waves: full result (solid) and result without helium (dotted).}
              \label{fig:lamcritslow}%
    \end{figure*}

We move to the slow magnetoacoustic waves and consider Equations~(\ref{eq:second1})--(\ref{eq:second3}). They can be combined to arrive at a rather involved equation for $v_\parallel'$ alone, namely
\begin{eqnarray}
&&\left[ \frac{\partial}{\partial t} - \left( \gamma -1 \right)\left( \tilde{\kappa}\frac{\partial^2}{\partial x^2} -\omega_T \right) \right] \left( \frac{\partial}{\partial t} - \frac{4\tilde{\zeta}_{\rm n}}{3} \frac{\partial^2}{\partial x^2}  \right) \frac{\partial v_\parallel'}{\partial t} \nonumber \\
&&- \cs^2 \frac{\partial^3 v_\parallel'}{\partial x^2 \partial t} + \left( \gamma -1 \right)\left[ \tilde{\kappa}\frac{\partial^2}{\partial x^2} -\frac{\cs^2}{\gamma}\left( \omega_T - \omega_\rho \right) \right]\frac{\partial^2 v_\parallel'}{\partial x^2} =  \nonumber \\
&&\frac{1}{2\mu\rho_0}\left[ \frac{\partial}{\partial t} - \left( \gamma -1 \right)\left( \tilde{\kappa}\frac{\partial^2}{\partial x^2} -\omega_T \right) \right] \frac{\partial^2 B_\perp'^2}{\partial x \partial t} \nonumber \\
&&- \left( \gamma -1 \right)\frac{\partial^2 }{\partial x \partial t} \left[ \frac{\etaCe}{\mu\rho_0}\left(\frac{\partial {B_\perp'}}{\partial x}\right)^2  + \tilde{\zeta}_{\rm n} \left(\frac{\partial {v_\perp'}}{\partial x}\right)^2  \right], \label{eq:slowlintot}
\end{eqnarray}
where $\cs^2 = \gamma p_0 /\rho_0$ is the  sound velocity squared and $\tilde\kappa$, $\omega_\rho$, and $\omega_T$ are defined as
\begin{eqnarray}
{\tilde \kappa} &\equiv & \frac{T_0}{p_0} \left( \kappa_{\rm e,\parallel} + \kappa_{\rm n}\right), \\
\omega_\rho &\equiv & \frac{\rho_0^2}{p_0} L_\rho,  \\
\omega_T &\equiv & \frac{\rho_0 T_0}{p_0} L_T.
\end{eqnarray}
 The general solution to  Equation~(\ref{eq:slowlintot}) is the sum of the  solution of the related homogeneous equation and the particular solution of the inhomogeneous equation. On the one hand, the solutions of the homogeneous equation  are uncoupled slow magnetoacoustic waves.  The uncoupled slow waves do not interact with the Alfv\'en waves studied in Section~\ref{sec:firstorder}. On the other hand,  the particular solution of the inhomogeneous equation  represents the  generation of slow magnetoacoustic waves due to the nonlinear coupling with the Alfv\'en waves. We note that the right-hand side of Equation~(\ref{eq:slowlintot}) involves terms with $B_\perp'$ and $v_\perp'$. These terms are associated with the Alfv\'en waves and are responsible for nonlinear driving slow  waves.  Here, we focus on the  solution of the  homogeneous version of Equation~(\ref{eq:slowlintot}) that represents uncoupled slow waves. The nonlinear driving of slow waves due to the coupling with the Alfv\'en waves is briefly addressed later in Appendix~\ref{sec:nonlienar}.  
 
 We perform  a Fourier analysis and write $v_\parallel'$ as
\begin{equation}
v_\parallel' = \sum_{k} \mathcal{V}_{\parallel,k} \exp \left[ i k x - i \omega(k) t \right], \label{eq:fourierslow}
\end{equation}
where $\mathcal{V}_{\parallel,k}$ is the amplitude of the $k$-th Fourier mode of $v_\parallel'$ and $k$ and $\omega(k)$ denote again the wavenumber and angular frequency. We obtain the dispersion relation from the  homogeneous version of Equation~(\ref{eq:slowlintot}), namely
\begin{equation}
\omega^3(k) + i a_2 \omega^2(k) - a_1 \omega(k) - i a_0 = 0, \label{eq:reldisperslow}
\end{equation}
with
\begin{eqnarray}
a_2 &=&  \frac{4\tilde{\zeta}_{\rm n}}{3} k^2 + \left( \gamma-1 \right)\left( \tilde{\kappa} k^2 + \omega_T \right),\\
a_1 &=&   k^2 \left( \cs^2 +  \left( \gamma-1 \right)\left( \tilde{\kappa} k^2 + \omega_T \right)\frac{4\tilde{\zeta}_{\rm n}}{3} \right), \\
a_0 &=& \left( \gamma-1 \right)\left( \tilde{\kappa} k^2 + \frac{\cs^2}{\gamma} \left( \omega_T - \omega_\rho \right) \right) k^2.
\end{eqnarray}
Equation~(\ref{eq:reldisperslow}) is a third-order polynomial and is the dispersion relation of parallel-propagating  slow magnetoacoustic waves, but it also describes thermal  modes \citep{1965ApJ...142..531F}. In general, slow waves and thermal modes are coupled when nonadiabatic effects are considered \citep[see, e.g.,][]{1994ApJ...435..482P,2004A&A...415..739C}. 

The thermal mode is the nonadiabatic version of the so-called entropy mode, which is related to nonpropagating perturbations of density in a plasma \citep[][]{2004prma.book.....G}. Following the same method as in \citet{2012A&A...540A...7S}, an approximate solution for the thermal mode in the case of weak damping can be found by neglecting terms of $\mathcal{O}\left(\omega^2\right)$ in Equation~(\ref{eq:reldisperslow}). By doing so, we find
\begin{equation}
\omega(k) \approx - i \frac{ \left( \gamma-1 \right)\left( \tilde{\kappa} k^2 + \frac{\cs^2}{\gamma} \left( \omega_T - \omega_\rho \right) \right)}{\cs^2 +  \left( \gamma-1 \right)\left( \tilde{\kappa} k^2 + \omega_T \right)\frac{4\tilde{\zeta}_{\rm n}}{3}}. \label{eq:thermal}
\end{equation}
The approximate thermal mode frequency given in Equation~(\ref{eq:thermal}) is purely imaginary. This means that thermal modes are always nonpropagating regardless of the spatial scale of their perturbations. For chromospheric conditions, the thermal mode imaginary part of the frequency is always negative, so that the thermal mode is a purely damped solution \citep[see, e.g.,][]{2012A&A...540A...7S}.  Growing modes related to thermal instabilities would require a positive imaginary part of the  frequency. Therefore, according to Equation~(\ref{eq:thermal}) thermal instability does not take place in the chromospheric plasma. Thermal instability requires higher temperatures and is though to occur in the coronal plasma  \citep[see details in][]{1965ApJ...142..531F}. For an optically thin plasma, the condition for thermal instability is that $\alpha < 0$, where $\alpha$ is the exponent of the temperature in Equation~(\ref{eq:radiationthin}). In the absence of thermal conduction and radiative losses, Equation~(\ref{eq:thermal}) gives $\omega(k) = 0$. This zero-frequency solution is the classic entropy mode in an adiabatic plasma \citep[see][]{2004prma.book.....G}. We shall not explore thermal modes further in this paper.

We go back to the study of slow waves. The exact analytic solution to Equation~(\ref{eq:reldisperslow}) is too complicated to obtain a simple expression of the critical dissipation lengthscale of slow waves. However, we can apply the method of \citet{2013ApJ...767..171S,2013ApJS..209...16S} and use the polynomial discriminant of Equation~(\ref{eq:reldisperslow}) to obtain some useful information. To do so, we perform the change of variable $\omega(k) = i s(k)$ in Equation~(\ref{eq:reldisperslow}) and remove the common factor $-i$. Then  a cubic equation in $s$ with real coefficients is obtained, namely
\begin{equation}
s(k)^3 + a_2 s(k)^2 + a_1 s(k) + a_0 =0
\end{equation}
Subsequently, we compute the polynomial discriminant, $\Delta$, of the cubic equation using the standard formula as
\begin{equation}
\Delta = a_2^2 a_1^2 - 4 a_1^3 - 4 a_2^3 a_0 - 27 a_0^2 + 18 a_2 a_1 a_0. 
\end{equation}
The full developed expression of the discriminant is not given here for the sake of simplicity. The discriminant has the general property that it is zero when the original polynomial has a multiple root. This is very convenient, since a multiple zero of Equation~(\ref{eq:reldisperslow}) occurs precisely when the slow wave suffers a cutoff \citep[see an explanation for the reason of this result in][]{2013ApJ...767..171S,2013ApJS..209...16S}. As a function of height we numerically determine  the real values of $k$, i.e., the cutoff wavenumbers, which cause the discriminant to vanish. We find that the discriminant has only one real root, namely $k_s$, which defines the  slow wave critical dissipation lengthscale as $\lambda_s = 2\pi/k_s$. The numerically obtained critical dissipation lengthscale of  slow waves is plotted in Figure~\ref{fig:lamcritslow}(a) as a function of height. The lengthscale varies several orders of magnitude from $\sim 10^{-4}$~m in the lower chromosphere to $\sim 10$~m in the upper chromosphere. Thus, the critical dissipation lengthscale for slow waves is significantly shorter than that obtained for Alfv\'en waves in Section~\ref{sec:firstorder}. Slow waves require very short spatial scales to be efficiently damped. Another difference between Alfv\'en waves and slow waves is that the slow wave critical dissipation  lengthscale is unaffected by the value of the magnetic field strength.

{ We note that the slow wave critical dissipation lengthscale reaches very short values in the lower chromosphere. These extremely short lengthscales are probably stretching the assumptions behind the  theory used in this paper. Such  short lengthscales are  of the same order of or shorter than the mean free path of particles between collisions. Therefore, the single-fluid MHD approach may be compromised for such small scales in the lower chromosphere. Multi-fluid or kinetic approaches would be more convenient instead. Readers must be aware that the results obtained here for the slow wave critical dissipation lengthscale in the lower part of the chromosphere should be interpreted with caution.}

In order to gain physical insight on the mechanisms responsible for the slow wave cutoff,  now we compute the critical dissipation lengthscale when only one specific damping mechanism is retained. In the case of radiative losses, we find that this mechanism cannot produce the cutoff of the slow wave. In other words, the slow wave never becomes critically damped because of the effect of radiative losses alone. The impact of radiative losses on the damping of slow waves in the chromosphere is therefore negligible. Conversely, the critical dissipation lengthscale is very similar to the full result when either thermal conduction or viscosity are considered alone (these results are overplotted in Figure~\ref{fig:lamcritslow}(a)). The two mechanisms are equally important and both should be considered to explain the slow wave  dissipation.

Taking advantage of the numerical result displayed in Figure~\ref{fig:lamcritslow}(a), we can try to find an analytic approximation of the slow wave critical dissipation lengthscale. The process is as follows. First, we  decouple the slow wave from the thermal mode by roughly neglecting the independent term from Equation~(\ref{eq:reldisperslow}). Then, we consider the numerical result that radiative losses are unimportant, so that we are allowed to neglect the terms with $\omega_T$ and $\omega_\rho$. This gives an approximate solution for the slow wave frequency as 
\begin{eqnarray}
\omega(k) &\approx & \pm k \cs \left[ 1  - \frac{k^2}{4\cs^2} \left( \frac{4\tilde{\zeta}_{\rm n}}{3} -  \left( \gamma-1 \right)\tilde{\kappa} \right)^{2} \right]^{1/2} \nonumber \\
&&- i \frac{k^2}{2} \left( \frac{4\tilde{\zeta}_{\rm n}}{3} +  \left( \gamma-1 \right) \tilde{\kappa}  \right), \label{eq:omegaslow}
\end{eqnarray}
where again the $+$ and $-$ signs in front of the real part of the frequency correspond to upward and downward propagating waves, respectively. We use  Equation~(\ref{eq:omegaslow}) to compute quality factor, $Q(k)$, namely
\begin{equation}
Q(k) \approx \left\{ \begin{array}{lll}
\frac{\left| \frac{4\tilde{\zeta}_{\rm n}}{3} - \left( \gamma - 1 \right) \tilde{\kappa}\right|}{\frac{4\tilde{\zeta}_{\rm n}}{3} + \left( \gamma - 1 \right) \tilde{\kappa}}\frac{\sqrt{k_s^2 - k^2}}{2k}  & {\rm if} & k < k_s, \\
0 & {\rm if} & k \geq k_s,  
\end{array} \right.
\end{equation}
where $k_s$ is the approximate cutoff wavenumber for slow waves given by
\begin{equation}
k_s \approx \frac{2\cs}{\left| \frac{4\tilde{\zeta}_{\rm n}}{3} - \left( \gamma - 1 \right) \tilde{\kappa} \right|}. \label{eq:cutoffslow}
\end{equation}
The critical dissipation lengthscale defined by this  cutoff wavenumber is 
\begin{equation}
\lambda_s = \frac{2\pi}{k_s} \approx  \frac{\pi\left| \frac{4\tilde{\zeta}_{\rm n}}{3} - \left( \gamma - 1 \right) \tilde{\kappa} \right|}{\cs}. \label{eq:slowlength}
\end{equation}
We overplot Equation~(\ref{eq:slowlength}) in Figure~\ref{fig:lamcritslow}(a). An excellent agreement between the approximate lengthscale and the numerically computed lengthscale is obtained. In addition, we can compare Equation~(\ref{eq:slowlength}) with the expression for the Alfv\'en wave critical dissipation lengthscale (Equation~\ref{eq:alfvencritical}). We see that both equations are formally similar and that thermal conduction plays in slow waves the same role as Cowling's diffusion plays in Alfv\'en waves.

As done in the case of Alfv\'en waves, we have also investigated the specific effect of helium on the critical dissipation lengthscale for slow magnetoacoustic waves. Figure~\ref{fig:lamcritslow}(b) shows the slow wave critical dissipation lengthscale  when helium is taken into account and when the effect of helium is omitted from the dissipation mechanisms. We obtain that the influence of helium is negligible except at large heights. As already discussed, the reason for this result is that helium remains mostly neutral at all heights in the chromosphere, while hydrogen gets fully ionized at large heights. Therefore, at those large heights there is a significant amount of neutral helium that efficiently contributes to both thermal conduction and viscosity.

\section{DISCUSSION}
\label{sec:discussion}

Strongly damped waves are good candidates to produce significant heating of the chromospheric plasma via conversion of wave energy into thermal energy. Thus, it is important to know the physical processes involved in the damping and the spatial scales at which these processes work. We have investigated the physical mechanisms that may be able to efficiently dissipate MHD wave energy in the chromosphere and have computed the spatial scales associated to critical damping (wave cutoffs).

By comparing the results of Alfv\'en waves (Figure~\ref{fig:lamcrit}) and slow magnetoacoustic waves (Figure~\ref{fig:lamcritslow}) we notice several important implications.  For instance, the physical mechanisms responsible for the occurrence of the cutoffs are different for the two types of MHD waves. The Alfv\'en wave cutoff owes its existence to the effect of Cowling's (Ohmic+ambipolar) diffusion, while viscosity plays a much less important role. For practical purposes the impact of viscosity on Alfv\'en waves can be safely neglected. In addition,  neither thermal conduction nor radiative losses affect the Alfv\'en wave damping at all, owing to the incompressible nature of Alfv\'en waves. Conversely, the  slow magnetoacoustic wave cutoff is caused by the combined effects of viscosity and thermal conduction, with the two mechanisms playing an equally important role at all heights in the chromosphere. Since slow waves are unaffected by the magnetic field Cowling's  diffusion does not play a role in the cutoff of slow waves, and radiative losses are again irrelevant. Among all the considered damping mechanisms, radiative losses is the only mechanism that can be safely neglected on the case of both Alfv\'en and slow waves. These results are consistent with previous estimations of the efficiency of various damping mechanisms in the chromosphere  \citep{2004A&A...422.1073K,2006AdSpR..37..447K}.

 \begin{figure}
   \centering
  \includegraphics[width=0.99\columnwidth]{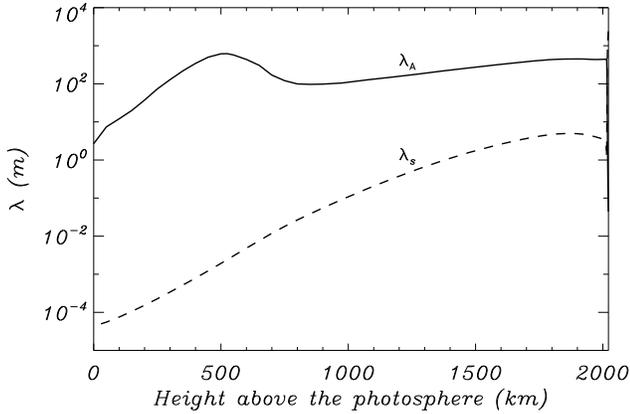}
   \caption{Critical dissipation lengthscale of Alfv\'en waves (solid line) as a function of height above the solar photosphere for a magnetic field strength that varies according to Equation~(\ref{eq:b0real}) with  $B_{\rm ph}=2$~kG. The critical dissipation lengthscale of slow waves (dashed line) is also plotted for comparison.}
              \label{fig:realb}%
    \end{figure}

Another relevant result is that the values of the critical dissipation lengthscales are rather different for Alfv\'en and slow magnetoacoustic waves. In the case of Alfv\'en waves, we find that the critical dissipation lengthscale is very sensitive to the magnetic field strength (see Figure~\ref{fig:lamcrit}(a)). In the low chromosphere the stronger the field, the smaller the dissipation lengthscale. Conversely, in the middle and high chromosphere the stronger the field, the larger the dissipation lengthscale. Because of the simplicity of our 1D model, we were restricted to consider straight and constant magnetic fields. In reality the chromospheric magnetic field is neither straight nor constant but composed of flux tubes that expand from the photosphere to the corona, with the magnetic field strength varying from strengths of kG in the photosphere to strengths of few G in the corona. To incorporate the effect of a magnetic field  that realistically varies with height, let us assume that the chromospheric magnetic field strength can be roughly approximated by the semi-empirical formula used by \citet{2006A&A...450..805L}, namely
\begin{equation}
B = B_{\rm ph} \left( \frac{\rho}{\rho_{\rm ph}} \right)^{0.3}, \label{eq:b0real}
\end{equation}
where $B_{\rm ph}$ and $\rho_{\rm ph}$ are the magnetic field strength and density at the photosphere. We compute the Alfv\'en wave critical dissipation lengthscale according to this prescription for the magnetic field strength. This result is displayed in Figure~\ref{fig:realb}. For this realistically varying field strength, the Alfv\'en wave critical dissipation lengthscale is roughly constrained in between 10~m (low chromosphere) and 1~km (medium and heigh chromosphere). We should note that the Alfv\'en wave critical dissipation lengthscale obtained for a varying field strength is roughly of the same order of magnitude as that obtained by \citet[see their Figure~3]{2015A&A...573A..79S}, although in that paper the effect of helium was ignored. Conversely, the slow magnetoacoustic wave critical lengthscale is unaffected by the magnetic field strength. For comparison purposes, we have overplotted in Figure~\ref{fig:realb} the  slow wave critical lengthscale that was already shown in Figure~\ref{fig:lamcritslow}(a). We see that the slow wave critical lengthscale is { several orders of magnitude shorter than that} obtained for Alfv\'en waves. This comparison informs us that the dissipation mechanisms working in the chromospheric plasma are more efficient in dissipating magnetic energy (associated to Alfv\'en waves) than acoustic energy (associated to slow magnetoaocustic waves), because the efficient dissipation of slow waves requires shorter spatial scales.

As mentioned in the Introduction, the observational evidence of MHD waves in the chromosphere is overwhelming. However, the fact that waves are observed does not automatically imply that those waves are actually contributing in heating the plasma. According to the results of this work, only the waves with the appropriate spatial scales (wavelengths) may be able to efficiently damp and deposit their energy in the chromospheric medium. These spatial scales,  below 1~km for Alfv\'en waves and even smaller for magnetoacoustic waves, are unresolved by current space-based and ground-based telescopes. In this direction,  future instruments operating at very high temporal and spatial resolutions like, e.g., the Atacama Large Millimeter/submillimeter Array (ALMA), may be necessary to observe those very short spatial scales and so to understand how MHD waves actually contribute to chromospheric plasma heating \citep{2015arXiv150406887W}.

\section{CONCLUDING REMARKS}
\label{sec:conclusions}

To conclude with, it is worth mentioning the possible impact of the simplifications used in this work. We have defined the critical dissipation lengthscales as the spatial scales at which strict wave cutoffs occur. A strict cutoff is mathematically defined so that the wave quality factor vanishes, i.e., $Q=0$. However, there are several physical effects not considered in the present investigation that could remove the mathematically strict wave cutoffs. \citet{2012A&A...544A.143Z}, and later \citet{2015A&A...573A..79S}, showed that Hall's term in the induction equation has the effect of removing the strict cutoffs. As explained in \citet{2012A&A...544A.143Z}, in the presence of Hall's term the waves do not suffer a strict cutoff when the wavenumber exceeds the critical value. Instead, the quality factor is nonzero but is always constrained in the interval $0 < Q < 1/2$. This is the interval corresponding to overdamped waves. In connection to the amount of wave energy that can be deposited in the plasma, the fact that the waves have mathematically strict cut-offs ($Q = 0$) or are overdamped ($0< Q < 1/2$) makes no practical difference. In the case of overdamped waves, the imaginary part of the frequency (related to damping) is  larger than the real part of the frequency (related to propagation). Therefore, as happens in a strict cutoff scenario, an overdamped MHD wave is unable to propagate, and so it cannot transport its energy away from the chromosphere \citep{2015A&A...573A..79S}. Hence, the actual condition for wave energy to be efficiently dissipated in the chromosphere should be $Q < 1/2$, which is less restrictive than the condition $Q=0$ imposed here. In both cases, the critical lengthscales are very similar. Another effect that can remove the strict cutoffs is the consideration of electron inertia terms in the multi-fluid description of the plasma \citep{2012A&A...544A.143Z,2015A&A...573A..79S}. As in the case of Hall's term,  electron inertia terms replace the strict cutoff by overdamping, but such a result has no practical implications concerning wave heating.  

We have assumed that chromospheric plasma dynamics can be studied under the single-fluid MHD approximation. Multi-fluid effects on the wave cutoffs have been investigated in detail by \citet{2013ApJ...767..171S,2013ApJS..209...16S,2015A&A...573A..79S}. They showed that wave cutoffs consistently occur in both single-fluid  and  multi-fluid theories. However, in the multi-fluid description it is found that waves may be able to propagate again when the lengthscale is further reduced and becomes smaller than a {\em second} critical value. As explained by \citet{2013ApJ...767..171S}, this second critical lengthscale would correspond to the spatial scale at which neutrals decouple from ions, and so the single-fluid approximation breaks down. Therefore, the existence of the second critical lengthscale is not captured by the single-fluid  description of the plasma. This second critical lengthscale  could be so small for chromospheric conditions that the fluid treatment of the plasma may become compromised \citep{2015A&A...573A..79S}. { Also, a more appropriate treatment of the slow wave critical dissipation in the lower chromosphere would require multi-fluid or kinetic approaches because of the extremely short spatial scales obtained.}

 We have considered a static one-dimensional model of the chromosphere, { which allows us} to analytically investigate the damping of Alfv\'en and slow MHD waves in the limit of small amplitudes and for spatial scales satisfying the local approximation.  The assumption of local analysis is fully justified in view that the critical dissipation lengthscales of both Alfv\'en and slow waves are much smaller than the pressure scale height in the chromosphere ($\sim 300$~km). Nevertheless, it should be acknowledged that a static model is a dramatic simplification because it misses the dynamic behavior of the chromosphere seen in high-resolution observations and reproduced by advanced numerical simulations \citep[e.g.,][]{2012ApJ...753..161M}. A time-dependent background should be necessary to account for the time-varying dynamics of the actual chromosphere. In addition, because of the linear, small-amplitude regime studied here, our approximate analysis is able to describe wave damping but misses the changes in the background due to wave energy deposition in the plasma. In order to properly overcome the limitations of the present analytic study, this paper needs to be extended in the future by considering fully nonlinear numerical simulations of MHD wave excitation and dissipation, including a time-dependent dynamic background and the self-consistent changes in the plasma due to  wave energy dissipation. This is an interesting task and will be tackled in a forthcoming work.

 \acknowledgements{{ We thank the referee for his/her constructive report.} We acknowledge support from MINECO and FEDER funds through project AYA2011-22846.  RS also acknowledges support from MINECO through a `Juan de la Cierva' grant, from MECD through project CEF11-0012, and from the `Vicerectorat d'Investigaci\'o i Postgrau' of the UIB. RS and JLB acknowledge discussion within the ISSI team on `Partially Ionized Plasmas in Astrophysics (PIPA)' and thank ISSI for their support. }

\bibliographystyle{apj} 
\bibliography{refs}

\begin{appendix}

\section{EFFECT OF DISSIPATION ON THE  COUPLING OF ALFV\'EN  AND SLOW WAVES}
\label{sec:nonlienar}

Here we briefly investigate the slow waves that are nonlinearly driven because of the presence of Alfv\'en waves. To do so, we must explore the solutions to the full Equation~(\ref{eq:slowlintot}), taking into account the inhomogeneous driving term on the right-hand side. We are mostly interested on the impact of the Alfv\'en wave dissipation on the process of mode coupling. Therefore, we shall focus on this particular aspect of the coupling between the two waves. A detailed analysis of the full evolution of nonlinear waves is left for future numerical studies.

The analytic study of the inhomogeneous Equation~(\ref{eq:slowlintot}) is difficult due to the complexity of the equation. In order to make further  progress and to study the wave coupling analytically, for simplicity we shall omit all dissipative mechanisms from Equation~(\ref{eq:slowlintot}). By doing so, we neglect the effect of dissipation on the nonlinearly driven slow waves. However, the effect of dissipation on the primary Alfv\'en wave is still considered at full. Under this approximation, Equation~(\ref{eq:slowlintot}) simplifies to
\begin{equation}
\frac{\partial^2 v_\parallel'}{\partial t^2} -  \cs^2 \frac{\partial^2 v_\parallel'}{\partial x^2}  = - \frac{1}{2\mu\rho_0}\frac{\partial^2 B_\perp'^2}{\partial x \partial t}. \label{eq:slowlin}
\end{equation}

To study the nonlinear coupling between the primary Alfv\'en wave and the nonlinearly driven slow wave, we  write $B_\perp'$ { in the form of} Equation~(\ref{eq:fourierby}). In turn, we write $v_\parallel'$ as in Equation~(\ref{eq:fourierslow}), but now we use $K$ and $\Omega$ to denote the wavenumber and frequency of slow waves. We use this different notation to distinguish the wavenumber and frequency of slow waves from the wavenumber and frequency of Alfv\'en waves, namely $k$ and $\omega$.  First, we substitute the expressions of $B_\perp'$ and $v_\parallel'$ into Equation~(\ref{eq:slowlin}) and obtain the relation between the wavenumbers and frequencies of the primary Alfv\'en wave and the nonlinearly driven slow wave, namely $K=2k$ and $\Omega=2\omega$. Therefore, we recover the typical result that the frequency and wavenumber of the nonlinearly generated slow mode are twice the frequency and wavenumber of the primary Alfv\'en mode. We shall use these relations to write all the following expressions in terms of $k$ and $\omega$. Next,  we find the relation between the  amplitudes of $B_\perp'$ and $v_\parallel'$ as
\begin{equation}
\mathcal{V}_{\parallel,k} = \left| \frac{k\omega(k)}{\omega^2(k) - k^2\cs^2 } \right| \frac{\mathcal{B}^2_k}{2\mu\rho}, \label{eq:relvparab}
\end{equation}
where $\omega(k)$ is the Alfv\'en wave frequency given by Equation~(\ref{eq:omegaimp}). 

\begin{figure*}
   \centering
  \includegraphics[width=0.45\columnwidth]{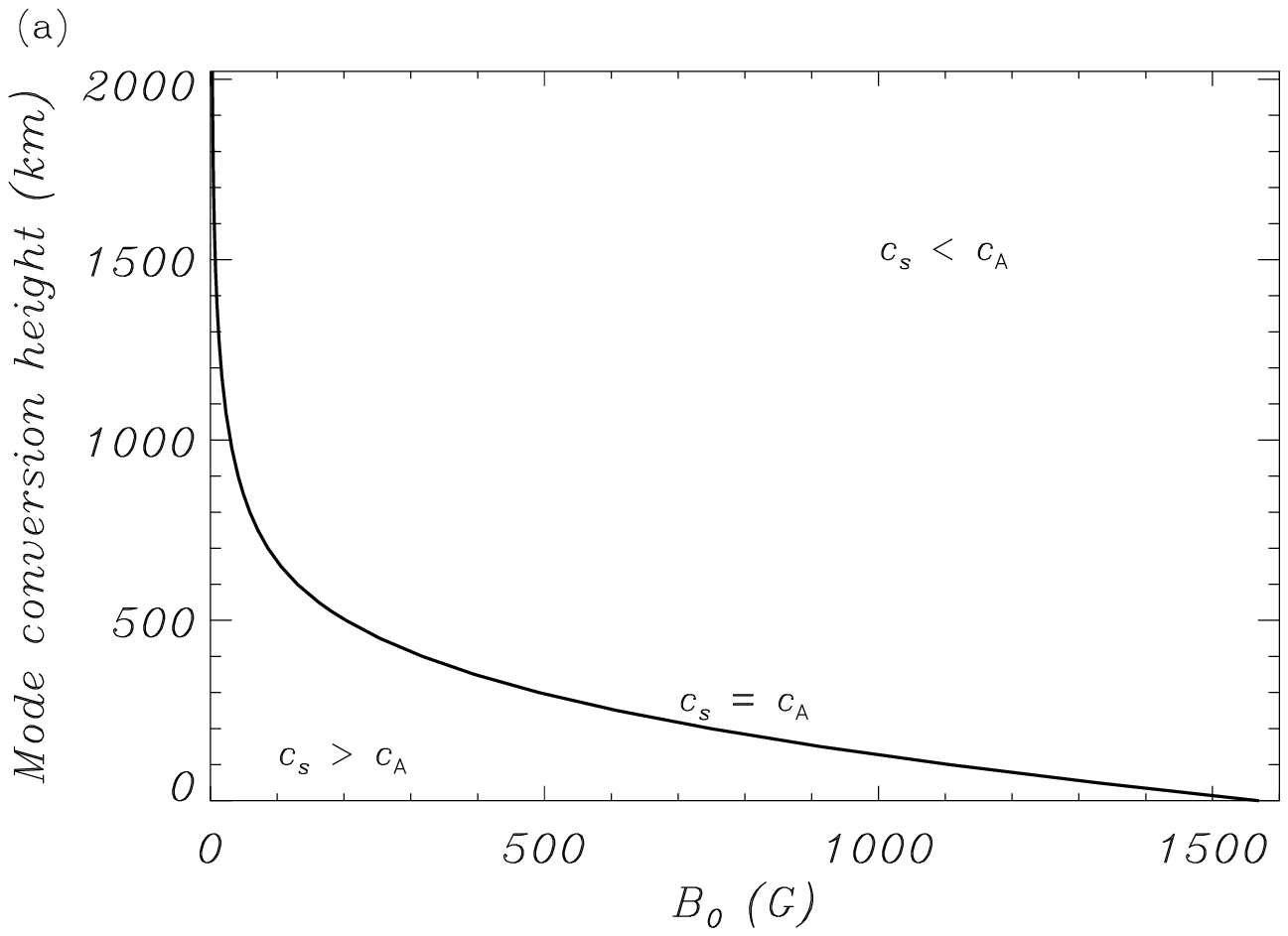}
   \includegraphics[width=0.45\columnwidth]{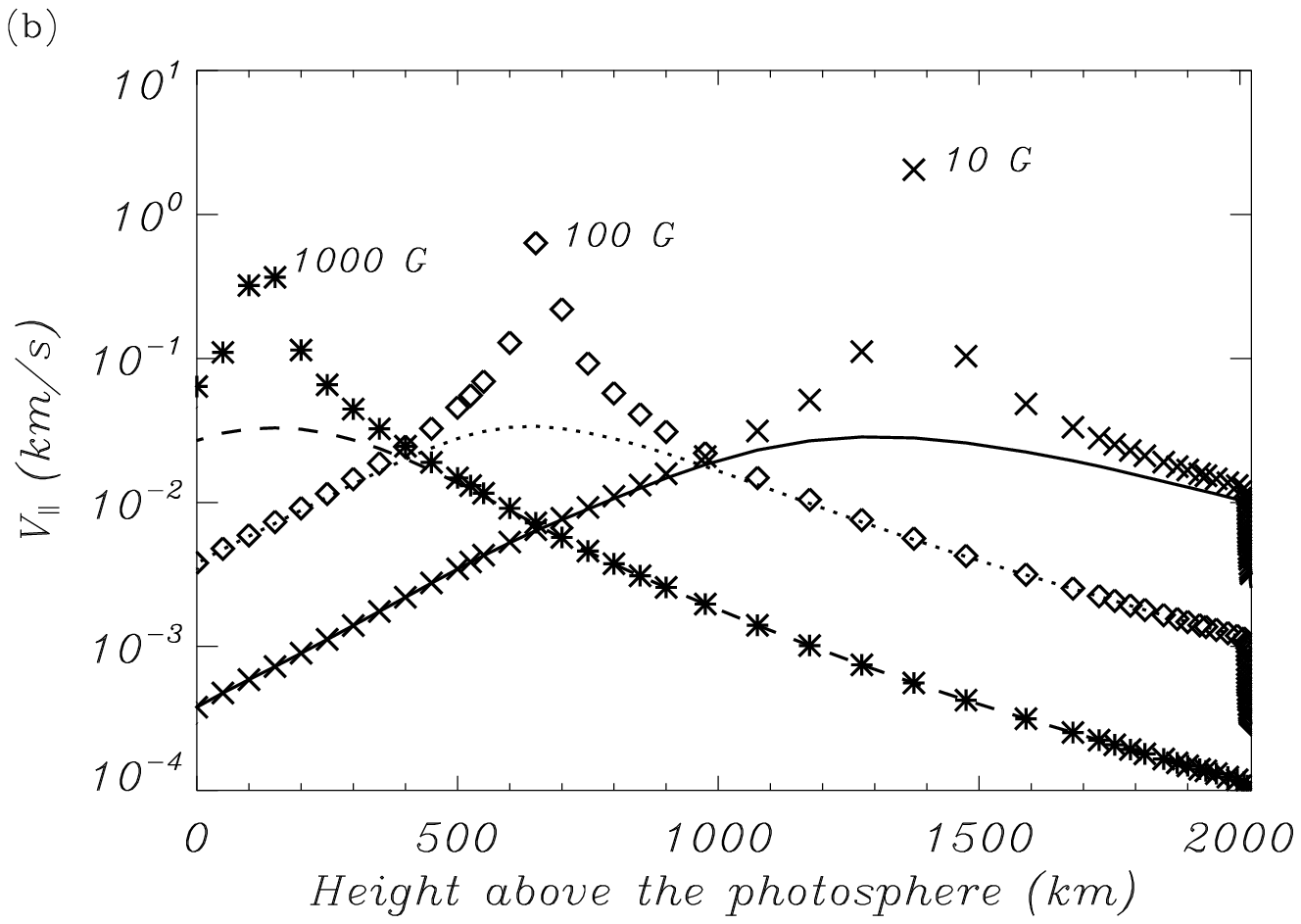}
   \caption{(a) Mode conversion height in the chromosphere as a function of the magnetic field strength in the dissipationless limit ($k\ll k_{\rm A}$). (b) Amplitude of the longitudinal component of velocity, $ \mathcal{V}_{\parallel,k}$, as a function of height in the chromosphere. Symbols denote results in the dissipationless limit ($k\ll k_{\rm A}$) and lines are the results in the limit of total dissipation ($k\to k_{\rm A}$). Three different values of the magnetic field strength are considered: 10~G (solid line and crosses), 100~G (dotted line and diamonds), and 1000~G (dashed line and asterisks). In all cases, we used $ \mathcal{V}_{\perp,k} = 1$~km/s as reference.}
              \label{fig:conversion}%
    \end{figure*}

It is more useful to write the relation between the amplitudes of two components of velocity, $v_\perp'$ and $v_\parallel'$. To do so, first we must find the relation  between  $v_\perp'$ and $B_\perp'$. Hence, we write
\begin{equation}
v_\perp' = \sum_k \mathcal{V}_{\perp,k} \exp\left[ i k x - i\omega(k) t  + i \varphi_0(k) \right],
\end{equation}
 where $\mathcal{V}_{\perp,k}$  is the amplitude of the Fourier mode of $v_\perp'$ with wavenumber $k$, and $\varphi_0(k)$ accounts for  phase differences between $v_\perp'$ and $B_\perp'$. If $k\leq k_{\rm A}$ the relation between $\mathcal{B}_k$ and $\mathcal{V}_{\perp,k}$ can be cast as
\begin{equation}
\mathcal{V}_{\perp,k} = \frac{\ca}{B_0}  \mathcal{B}_k , \label{eq:relationamp}
\end{equation}
while $\varphi_0(k)$ is  given by
\begin{equation}
\varphi_0(k) =   \pi \pm \arctan \left( \frac{k}{\sqrt{k_{\rm A}^2-k^2}} \right),
\end{equation}
where, as before, the $+$ and $-$ signs correspond to upward and downward propagating waves, respectively. Therefore, the relation between the amplitudes of $v_\perp'$ and $B_\perp'$ is constant, while the phase shift   depends on $k$. The phase shift is $\pi$ for $k\ll k_{\rm A}$ and $ \mp \pi/2$ when $k\to k_{\rm A}$.  Now, we use Equations~(\ref{eq:relvparab}) and (\ref{eq:relationamp}) to write the relation between the amplitudes of two components of velocity as
\begin{equation}
\mathcal{V}_{\parallel,k} = \left| \frac{k\omega(k)}{\omega^2(k) - k^2\cs^2 } \right| \frac{\mathcal{V}_{\perp,k}^2}{2}.
\end{equation}
In general, the relation of amplitudes of the two velocity components is a function of $k$. When $k\ll k_{\rm A}$, the damping of Alfv\'en waves is very weak and we can approximate the Alfv\'en wave frequency by $\omega(k)\approx \pm k \ca$. Then,  the relation of amplitudes  of the two velocity components simplifies to
\begin{equation}
\mathcal{V}_{\parallel,k} \approx \frac{\ca}{2 \left|\ca^2 - \cs^2  \right| } \mathcal{V}_{\perp,k}^2, \label{eq:coupling1}
\end{equation}
which is independent of $k$. Equation~(\ref{eq:coupling1}) shows that the parallel velocity component associated to the nonlinearly driven slow waves diverges when $\ca^2 = \cs^2$. This is the condition of mode conversion from Alfv\'en waves to slow waves.  Since $\ca$ is a function of the magnetic field strength, the specific height in the chromosphere at which the condition $\ca^2 = \cs^2$ is satisfied depends on the value of $B_0$. Figure~\ref{fig:conversion}(a) shows the mode conversion height as a function of the magnetic field strength. For weak fields, the condition $\ca^2 = \cs^2$ happens in the medium and/or upper chromosphere, while the mode conversion height moves to the low chromosphere when strong magnetic fields are considered. We have chosen $ \mathcal{V}_{\perp,k} = 1$~km/s as a reference value for the amplitude of the transverse component of velocity and have plotted in  Figure~\ref{fig:conversion}(b) the amplitude of the longitudinal component for three different values of the magnetic field strength. As predicted by Equation~(\ref{eq:coupling1}), the longitudinal component of velocity diverges at those specific heights where $\ca^2 = \cs^2$. It is at those heights that the { nonlinear driving} of slow waves is most effective. In the dissipationless limit ($k\ll k_{\rm A}$), energy from the Alfv\'en waves may be efficiently transferred to the slow waves at those specific heights.

Conversely, in the limit $k\to k_{\rm A}$ the Alfv\'en waves are fully damped and their frequency approximates by $\omega(k) \approx - i k^2 \left( \etaCe + \tilde{\zeta}_{\rm n} \right)/2$. In such a case, the relation of amplitudes  of the two velocity components tends to
\begin{equation}
\mathcal{V}_{\parallel,k} \approx \frac{1}{2} \frac{\frac{ \etaCe + \tilde{\zeta}_{\rm n}}{\left| \etaCe - \tilde{\zeta}_{\rm n}\right|}\ca}{\left( \frac{ \etaCe + \tilde{\zeta}_{\rm n}}{\etaCe - \tilde{\zeta}_{\rm n}} \right)^2 \ca^2 + \cs^2  } \mathcal{V}_{\perp,k}^2. \label{eq:coupling2}
\end{equation}
We have used  Equation~(\ref{eq:coupling2}) to overplot in Figure~\ref{fig:conversion}(b) the amplitude of the longitudinal component of velocity in the limit of strong diffusion of Alfv\'en waves ($k\to k_{\rm A}$). We see that the infinite amplitudes obtained in the difussionless case are replaced by finite amplitudes. To estimate the maximum amplitude of $\mathcal{V}_{\parallel,k}$, we set $\ca^2 = \cs^2$ and consider that the effect of viscosity is much less important than that of Cowling's diffusion for the damping of Alfv\'en waves. Hence we  neglect viscosity from Equation~(\ref{eq:coupling2}). Then, the equation becomes
\begin{equation}
\max\left(\mathcal{V}_{\parallel,k}\right) \approx \frac{1}{4\ca}  \mathcal{V}_{\perp,k}^2. \label{eq:coupling3}
\end{equation}
Equation~(\ref{eq:coupling3}) indicates that for small-amplitude Alfv\'en waves with $\mathcal{V}_{\perp,k} \ll \ca$, the maximum amplitude of the longitudinal component of velocity is necessarily very small in the limit of strong dissipation. 

The approximate results obtained here suggest that the nonlinear coupling of Alfv\'en and slow waves is affected by dissipation of the primary Alfv\'en wave. Dissipation may reduce the efficiency of Alfv\'en-to-slow mode conversion in the chromospheric layer where $\ca^2 = \cs^2$. However, these results are obtained under quite restrictive approximations and their validity should be checked by means of fully nonlinear numerical simulations. We leave this task for forthcoming works.

\end{appendix}

\end{document}